\documentclass[aps,prd,superscriptaddress,showkeys,showpacs,twocolumn,a4paper]{revtex4-1}
\usepackage{ulem}
\usepackage{color}
\usepackage{graphicx}
\usepackage{soul}
\usepackage{appendix}
\usepackage{epsfig}
\usepackage[usenames,dvipsnames,svgnames]{xcolor}
\definecolor{purple}{rgb}{0.3,0,0.9} % red blue
\definecolor{blue1}{rgb}{8, 24, 168}
\definecolor{darkteal}{HTML}{045D5D}
\usepackage[colorlinks=true,linkcolor= blue, citecolor = blue, urlcolor = purple ]{hyperref}
\usepackage{epstopdf}
\usepackage{amsmath,amssymb,amsthm,amsfonts} % For including math equations, theorems, symbols, etc
\usepackage{subfig}
\usepackage{comment}

\UseRawInputEncoding

\usepackage{float}

\newcommand{\be}{\begin{equation}}
\newcommand{\ee}{\end{equation}}
\newcommand{\ba}{\begin{eqnarray}}
\newcommand{\ea}{\end{eqnarray}}
\newcommand{\nn}{\nonumber\\}

\begin{document}

\title{Memory  effect on the heavy quark dynamics in
hot QCD matter}

\author{Jai Prakash}
%\email{jaiprakashaggrawal2@gmail.com}
\affiliation{State Key Laboratory of Dark Matter Physics, Shanghai Key Laboratory for Particle Physics and Cosmology,
Key Laboratory for Particle Astrophysics and Cosmology (MOE),
School of Physics and Astronomy, Shanghai Jiao Tong University, Shanghai 200240, China}

\author{Ling Hai Li }
%\email{jaiprakashaggrawal2@gmail.com}
\affiliation{State Key Laboratory of Dark Matter Physics, Shanghai Key Laboratory for Particle Physics and Cosmology,
Key Laboratory for Particle Astrophysics and Cosmology (MOE),
School of Physics and Astronomy, Shanghai Jiao Tong University, Shanghai 200240, China}

\author{Ying Shan Zhao}
%\email{jaiprakashaggrawal2@gmail.com}
\affiliation{State Key Laboratory of Dark Matter Physics, Shanghai Key Laboratory for Particle Physics and Cosmology,
Key Laboratory for Particle Astrophysics and Cosmology (MOE),
School of Physics and Astronomy, Shanghai Jiao Tong University, Shanghai 200240, China}

\author{Yifeng Sun}
\email{Correspond to: sunyfphy@sjtu.edu.cn}
\affiliation{State Key Laboratory of Dark Matter Physics, Shanghai Key Laboratory for Particle Physics and Cosmology,
Key Laboratory for Particle Astrophysics and Cosmology (MOE),
School of Physics and Astronomy, Shanghai Jiao Tong University, Shanghai 200240, China}

%\date{\today}

\begin{abstract}
We study the heavy quark dynamics in the presence of memory within the framework of a generalized Langevin equation. Time correlated thermal noise with power-law decay is generated by a fractional differential equation, formulated using the Caputo fractional derivative with order parameter $\nu$. The effect of memory is calculated through the momentum correlation, the time evolution of the average squared momentum, the average squared displacement, and the average kinetic energy. The effect of memory is further studied for the higher normalised central moments of the heavy quark transverse-momentum distribution.  The results indicate that time correlated thermal noise substantially influences heavy quark dynamics in the quark gluon plasma.
\end{abstract}

%%%%%%%%%%%%%%%%%%%%%%%%%%%%%%%%%%%%%%%%%%%%%%%%%%%%%%%%%%%%%%%%%%%%%%%


\maketitle

%%%%%%%%%%%%%%%%%%%%%%%%%%%%%%%%%%%%%%%%%%%%%%%%%%%%%%%%%%%%%%%%%%%%%%%

%%%%%%%%%%%%%%%%
\section{Introduction}
%%%%%%%%%%%%  new %%%%%%%%%%%%%
Ultra-relativistic heavy-ion collisions at the Relativistic Heavy-Ion Collider (RHIC) and the Large Hadron Collider (LHC) have predicted the formation of quark-gluon plasma (QGP), a deconfined state of strongly interacting matter in which quarks and gluons are no longer confined into hadrons \cite{STAR:2006vcp, Adams:2005dq, PHENIX:2004vcz, ALICE:2010khr, BRAHMS:2005gow}. This medium exists only for a short time, with an estimated lifetime of a few fm/c \cite{vanHees:2004gq, Rapp:2009my}. The study of the properties of the QGP remains a subject of considerable current interest. In this context, the heavy quarks (HQs), namely charm and beauty, are regarded as useful probes of the medium produced in high-ion collisions (HICs) ~\cite{Cao:2016gvr,Cao:2015hia,scardina2017estimating,He:2022ywp,Song:2015sfa, Andronic:2015wma, Dong:2019unq, Aarts:2016hap,Plumari:2017ntm, Gossiaux:2008jv, Prakash:2021lwt, Prakash:2023wbs, Prakash:2023hfj, Jamal:2023ncn,ZACCONE2024116483, Singh:2023smw, Kurian:2020orp, Mazumder:2011nj, PhysRevD.103.054030, Jamal:2021btg, Jamal:2020emj, Sun:2023adv, Plumari:2019hzp, Prakash_2024, du2023accelerated, Shaikh:2021lka, Kumar:2021goi,Sumit:2025ddb,PhysRevLett.130.231902,Das:2024vac,Chandra:2024ron,Debnath:2023zet,Jamal:2025gjy,Das:2024wht,Das:2022lqh,Sambataro:2025pop,Das:2025yxy,Dey:2025kqx,Minissale:2023dct,Oliva:2024rex,PhysRevC.84.044901,Bhattacharyya:2024hku}.  The HQs are produced predominantly through initial hard scatterings in the early stage of the HICs. Since their masses are much larger than the characteristic temperatures reached in HICs, their thermalization proceeds more slowly than that of the light partonic constituents of the medium. They therefore serve as effective probes of the system throughout its evolution, from the initial stage of the collision to hadronization.

The theoretical description of position and momentum evolution of the HQs is commonly formulated in terms of stochastic transport equations, in particular, the Langevin equation \cite{PhysRevC.86.034905, PhysRevC.93.014901, PhysRevC.84.064902,vanHees:2007me}. In the standard Langevin equation, the thermal noise is typically assumed to be Gaussian white noise \cite{Cao:2018ews, Rapp:2018qla, Prino:2016cni,He:2013zua,Xu:2018gux,He:2012df} and is therefore assumed to be uncorrelated in time. This approximation, however, may not provide a complete description \cite{122333,Zhang:2022fum} of the QGP medium.  If the stochastic force retains information about prior interactions between the HQs and the medium, the dynamics become non-Markovian and require a formulation that incorporates memory effects.  In the present work, this assumption is relaxed and the case of time correlated thermal noise is discussed.
 The relevance of such effects in stochastic modeling has been emphasized in several recent studies \cite{PhysRevLett.98.022301, PhysRevC.85.054906, Ruggieri:2019zos, Schuller:2019ega, PhysRevE.107.064131, PhysRevC.49.1693}, and non-Markovian dynamics have been studied in the nuclear fission process \cite{Gegechkori:2008ppw, Ivanyuk:2021tzw}.

Within QCD matter, time correlated thermal noise, often referred to as colored noise, has been incorporated into studies of baryon-number diffusion \cite{Kapusta:2014dja}, hydrodynamic fluctuations \cite{Murase:2013tma}, and the frequency-dependent electrical conductivity of hot QCD \cite{Hammelmann:2018ath}. In the context of the HQ dynamics in hot QCD matter, memory effects have been studied by modeling the thermal noise with time correlations. An exponentially decaying noise correlator was employed in Ref.~\cite{Ruggieri:2022kxv} to study its impact on the HQ momentum evolution and the nuclear modification factor, $R_{AA}$. In a complementary approach, Ref.~\cite{Pooja:2023gqt} considered long-tailed correlations with power-law decay and incorporated memory through a Riemann--Liouville (R--L) fractional integral of the noise.

In this context, the Caputo fractional derivative is particularly well explored \cite{guo2013numerics,li2013finite,miller1993introduction,CaputoMainardi1971}. By contrast, the R--L derivative \cite{carpinteri2014fractals} requires fractional-order initial conditions, which generally lack direct physical interpretation and complicate numerical implementation. In transport theory and stochastic dynamics, the Caputo formulation has therefore been adopted extensively because it yields solutions that are regular at the initial time and consistent with the expected long-time physical behavior \cite{Ciesielski2003}. Systematic comparisons presented in Refs.~\cite{Sandev2015} accordingly favor the Caputo definition for physical modelling, while Ref.~\cite{Fa2007} shows that Langevin equations formulated with the Caputo derivative produce correlation structures that are more physically consistent than those obtained from the R--L definition.

The present work employs the Caputo fractional derivative to construct time-correlated thermal noise with power-law decay and incorporates it into a generalized Langevin equation (GLE) framework for the dynamics of the HQs in the QGP medium. In this paper, the HQs are assumed to propagate through a thermalized QGP medium at a fixed temperature($T$). Within this formulation, the impact of memory is explored systematically through its effects on the HQ momentum correlation, the average squared momentum, the average squared displacement, the average kinetic energy, and the higher normalised central moments. This framework, therefore, provides a systematic basis for analysing non-Markovian effects generated by time-correlated thermal noise in the HQ dynamics in the QGP medium.

%%%%%%%%%%% %%%%%%%%%%%%%

The paper is organized as follows. Section~\ref{FORM} outlines the theoretical and numerical framework and details the implementation of power-law correlated processes. Section~\ref{res} presents the numerical results and analysis. A summary of the findings and concluding remarks is given in Section ~\ref{con}.

%%%  old version %%%%%%%%%%%%%%
\section{ FORMALISM}
\label{FORM}
\subsection{Time correlated thermal noise}
Long-tailed thermal noise with power-law time correlations is generated by the fractional stochastic equation, which is defined as follows,
\begin{align}\label{stochastic}
 \tau^{\nu}[ {^{C} D^\nu_{0+}}h(t)] = \eta(t),
\end{align}
\(^{C}D_{0+}^{\nu}\) represents the Caputo derivative operator \cite{lim2009modeling} and $\nu$ denotes the fractional order parameter, satisfying the condition \(q-1<\nu\leq q\) for \(q\in\mathbb{N}\) ($\mathbb{N}$ is a natural number). The parameter $\tau$, which carries the dimension of time, is introduced to maintain the dimensional consistency of the stochastic equation. Specifically, the factor $\tau^{\nu}$ compensates the time dimension associated with the fractional derivative in Eq.~\eqref{stochastic}, so that the stochastic process $h(t)$ remains dimensionless in this formulation. The thermal noise \(\eta(t)\) is assumed to be standard Gaussian noise and satisfies the correlation as follows,
\begin{align}\label{corr_g}
\langle\eta(t)\eta(u)\rangle=\tau\delta(t-u),
\end{align}
with $\langle\eta(t)\rangle=0$.
The explicit form of the Caputo fractional derivative in Eq.~\eqref{stochastic} is given by~\cite{10.1111/j.1365-246X.1967.tb02303.x},
 \begin{align}\label{Caputo}
 ^{C} D^\nu_{0+}u(t) = \frac{1}{\Gamma({q-\nu})}\int_0^t \frac{u^{(q)}(s)}{(t-s)^{1+\nu-q}}ds,   
\end{align}
 \(u_{(q)}\) denotes the \(q\)th derivative of \(u\), and \(\Gamma(\cdot)\) denotes the gamma function. To generate time-correlated thermal noise with power-law decay, the analysis is restricted to the range $0<\nu\le1$, so $q=1$. Accordingly, the Caputo derivative reduces to~\cite{Mainardi2007SubdiffusionEO,PhysRevLett.93.180603,Prakash:2024irm}
\begin{align}\label{subD}
 ^{C} D^\nu_{0+}u(t) = \frac{1}{\Gamma({1-\nu})}\int_0^t \frac{u^{(1)}(s)}{(t-s)^{\nu}}ds,    
\end{align}

%%%%%%%%%%%%%%%%%%%%
where $u^{(1)}(s)$ is the first derivative of $u(s)$. To obtain the subsequent analytical solution of Eq.~\eqref{stochastic}, it is useful to use the Laplace transform of the Caputo fractional derivative, namely~\cite{podlubnyacademic}
\begin{align}\label{laplace_h}
 \mathcal{L}\left[{}^{C} D^\nu_{0+}u(t)\right](z)
 =
 z^\nu\widehat{u}(z)
 -
 \sum_{k=0}^{m-1}z^{\nu-k-1}\left[u^{(k)}(0)\right],
\end{align}
where the notation $\widehat{(\,\cdot\,)}$ is used to denote the Laplace transform, \textit{i.e.}, $\widehat{u}(z)\equiv \mathcal{L}[u(t)](z)$. Applying the Laplace transform to Eq.~\eqref{stochastic},
\begin{align}
\tau^{\nu}\mathcal{L}\!\left[{}^{C} D^\nu_{0+}h(t)\right](z)=\widehat{\eta}(z),
\end{align}
and using Eq.~\eqref{laplace_h}, one obtains
\begin{align}
\tau^{\nu}\left[z^\nu \widehat{h}(z)-\sum_{k=0}^{m-1}z^{\nu-k-1}h^{(k)}(0)\right]=\widehat{\eta}(z),
\end{align}
here, $m$ denotes the integer satisfying $m-1<\nu\leq m$. In the present case, since $0<\nu\leq 1$, one has $m=1$, so that only the initial condition $h(0)$ contributes and the above equation reduces to
\begin{align}
\tau^{\nu}\left[z^\nu \widehat{h}(z)-z^{\nu-1}h(0)\right]=\widehat{\eta}(z),
\end{align}
from which one finally finds
\begin{align}\label{sublaplaceofh2}
\widehat{h}(z)=\frac{\widehat{\eta}(z)}{\tau^\nu z^\nu}+\frac{h(0)}{z}.
\end{align}
the inverse Laplace transform of Eq.~\eqref{sublaplaceofh2}, together with the initial condition \(h(0)=0\), yields
\begin{align}\label{h_evolution}
h(t) &= \frac{1}{\tau^{\nu}\,\Gamma(\nu)}\int_{0}^{t} (t-s)^{\nu-1}\,\eta(s)\,ds,
\qquad 0<\nu\le 1.
\end{align}
Using Eq.~\eqref{corr_g}, the two-point correlation function of the process \(h(t)\) can be simplified to

\begin{align}
\label{eq:corr_integral}
\langle h(t_1)h(t_2)\rangle
= \frac{\tau^{1-2\nu}}{\Gamma(\nu)^2}
\int_{0}^{t_<} (t_1-s)^{\nu-1}(t_2-s)^{\nu-1}\,ds,
\end{align}
which holds for all  \(t_1,t_2>0\). Here, the convenient shorthand
\begin{equation}
t_> \;:=\; \max(t_1,t_2),\qquad
t_< \;:=\; \min(t_1,t_2).
\end{equation}
Upon performing the substitution \(s=t_{<}u\) and subsequently applying the Euler--Gauss integral representation of the Gauss hypergeometric function, one obtains
\begin{align}
\label{eq:corr_closed}
\langle h(t_1)h(t_2)\rangle
= \frac{\tau^{1-2\nu}}{\nu\,\Gamma(\nu)^2}\;
t_>^{\nu-1}\,t_<^{\nu}\;
{}_2F_1\!\left(1-\nu,\,1;\,1+\nu;\,\frac{t_<}{t_>}\right).
\end{align}
For \(0<\nu\leq 1\) and \(|\frac{t_<}{t_>}|<1\), the Gauss hypergeometric function admits the convergent series representation as follows,
\begin{align}
\label{eq:corr_series}
{}_2F_1(1-\nu,1;1+\nu;\lambda)
= \sum_{n=0}^{\infty} \frac{(1-\nu)_r}{(1+\nu)_r}\,\lambda^r,
\qquad
\lambda=\frac{t_<}{t_>},
\end{align}
where \((x)_n=\Gamma(x+r)/\Gamma(x)\) denotes the Pochhammer symbol, with \((x)_0=1\). In the asymptotic regime of well-separated times considered here, \(t_>/t_<\sim 20\), corresponding to \(\lambda\sim 0.05\ll 1\), the series converges rapidly. Consequently, 
\begin{align}
{}_2F_1(1-\nu,1;1+\nu;\lambda)
= 1+\mathcal{O}(\lambda),
\end{align}
the hypergeometric factor thus contributes only a small correction and can be neglected. In the asymptotic regime $t_1 \gg t_2 > 0$,  the two-point correlator reduces to
\begin{align}\label{corr2}
\langle h(t_1)h(t_2)\rangle
\sim \frac{\tau^{\,1-2\nu}}{\nu\,\Gamma(\nu)^2}\;
t_1^{\,\nu-1}\,t_2^{\,\nu}.
\end{align}
For $0 < \nu \leq 1$, Eq. \eqref{corr2} is finite. Consequently, for fixed \(t_2\), the correlations of the process \(h(t)\) decrease as \(t_1\) becomes larger than \(t_2\), following a power-law decay. The power-law decay of these correlations implies that the \(h(t)\) exhibits long-tailed memory. It is further noted that the correlator is not a function of the time difference \(t_1-t_2\), but instead depends separately on \(t_1\) and \(t_2\).   With the analytical expression for the thermal-noise correlation function established, the discussion now proceeds to its numerical implementation.

%%%%%%%%%%%%%%%%%%%%%%%%%%%%%%

\subsection{NUMERICAL IMPLEMENTATION}

In this subsection, we describe the numerical implementation of the stochastic process \(h(t)\) defined in Eq.~\eqref{stochastic}. Since the present formulation involves a fractional differential operator acting on the noise process $h(t)$, standard stochastic calculus frameworks based on  It\^{o} or Stratonovich integration are not directly applicable. We emphasize that 
$h(t)$  is generated by a Caputo fractional  operator driven by standard white noise $\eta(t)$. The inapplicability of the standard It\^{o} framework arises because the fractional operator introduces a non-local convolution in time, which renders the resulting process $h(t)$ non-Markovian and prevents its representation as a semimartingale~\cite{rogers1997arbitrage}. For this reason,  the Caputo derivative is discretized directly using the L1 scheme, as detailed  below ( as detailed in Appendix ~\ref{AP}),

\begin{widetext}
\begin{align}\label{SubA_x}
\begin{cases}
  h(t_1)= h(t_0) + \sqrt\frac{\tau}{\Delta t}\tilde{\zeta}(t) k_1  \; &:\;n=1, \\
  h(t_n)= h(t_{n-1})+ \Bigg[\displaystyle \sqrt\frac{\tau}{\Delta t}\tilde{\zeta}(t)  
 -\sum_{j=1}^{n-1} a_j \left(h(t_{n-j})-h(t_{n-j-1}))\right)\Bigg]k_1 \; &:\;n \geq 2, 
 \end{cases}
\end{align}
\end{widetext}
 where  \(\Delta t\) denotes the time step, the coefficients {$a_j =\frac{((j+1)^{1-\nu}-j^{1-\nu})}{\Delta t^{\nu}\Gamma(2-\nu) }$} for $1 \leq j \leq n-1$  and $k_1=\frac{\Delta t^{\nu}\Gamma(2-\nu)}{\tau^{\nu}}$.
Numerical methods for fractional differential equations are generally classified as indirect or direct: indirect methods reformulate the time-fractional differential equation as an integro-differential equation, whereas direct methods approximate the time-fractional derivative itself. The present analysis follows the latter strategy and discretizes the fractional derivative directly, without transforming the differential equation into its integral form. This distinguishes 
the present approach from that of Ref.~\cite{Pooja:2023gqt}, where the memory 
was incorporated through an integral representation of the noise. The construction of the \(h(t)\) correlation ensures a nonvanishing two-time correlation function and thus provides a time correlated noise for the HQs. \(\eta(t)\) in Eq.~\eqref{stochastic} is rescaled according to
\begin{align}    
\eta(t)=\tilde{\zeta}(t)\sqrt{\frac{\tau}{\Delta t}},
\label{34}
\end{align}
since
\begin{align}
\delta(t-u)\rightarrow \frac{\delta_{t,u}}{\Delta t}.
\label{33}
\end{align}
Accordingly, \(\tilde{\zeta}(t)\)   is generated at each time step to satisfy
\begin{align}
\langle\tilde{\zeta}(t)\tilde{\zeta}(u)\rangle=\delta_{t,u},
\label{corr}
\end{align}
For the discretization convention, $t_0$ denotes the initial time and $n$ denotes the total number of time steps, so that the time at the $i^{\mathrm{th}}$ step is given by $t_i = t_0 + i\,\Delta t$, with $t = t_n$ at the $n^{\mathrm{th}}$ step. Once the colored thermal noise bath has been generated in Eq.~\eqref{SubA_x}, it is embedded into the discretized GLE for the HQ dynamics, as detailed in the following subsection. The numerical scheme for the coupled evolution of the process $h(t)$ and the HQ motion within the GLE framework is then presented in the subsequent subsection.

\subsection{Generalized Langevin equation}
\label{GLE}

The position and the momentum evolution of the particle are described by the GLE \cite{Lim2009ModelingSD,SANDEV20113627,SandevMetzlerTomovski,2014JMP....55b3301S},

\begin{align}\label{Langevin_x_rel}
&\frac{dx(t)}{dt}=\frac{p(t)}{E(t)},
\\
&\frac{dp(t)}{dt} = - \int_0^t \gamma (t,u) p(u) du+\xi(t),
\label{Langevin_p}
\end{align}
where $p(t)$ and $x(t)$ denote the momentum and the position of the HQs, respectively, and $E(t)$ is its energy. The HQ is subject to a deterministic dissipative force characterized by the memory kernel, \(\gamma(t,u)\), and a stochastic force \(\xi(t)\), whose properties are specified by,

\begin{align}
 &\langle\xi(t)\rangle=0,\\
&\langle\xi(t)\xi(t')\rangle=\frac{2\mathcal{D}f(t,t')}{\tau}\label{corr1}, 
\end{align}
where $\mathcal{D}$ is the diffusion coefficient of the HQs and $f$ is a dimensionless function defining the correlation of the noise; the factor $1/\tau$ in Eq. \eqref{corr1} is introduced to balance dimensions in the equation. In the Markovian limit, $f(t)/\tau = \delta(t)$. The stochastic force is related to \(h(t)\) by 

\begin{equation}\label{h_rel}
    \xi(t)= \sqrt{\frac{2\mathcal{D}}{\tau}} h(t),
\end{equation}
hence,
\begin{align}
    \label{corr3}&\langle \xi(t)\xi(t')\rangle=\frac{2\mathcal{D}}{\tau}\langle h(t)h(t')\rangle.
\end{align}
Comparison of Eqs.~\eqref{corr1} and~\eqref{corr3} yields

\begin{align}
    \label{corr4}&f(t',t)=\langle h(t)h(t')\rangle.
\end{align}
The stochastic GLE in terms of the colored noise, $h (t)$ is ~\cite{Das:2013kea, Ruggieri:2022kxv},

\begin{align}
\frac{dp(t)}{dt} = - \int_0^t \gamma (t,u) p(u) du+ \sqrt{\frac{{2\mathcal{D}}}{\tau}}h(t),
\label{Langevin_ph}
\end{align}
the dissipation kernel,  $\gamma(t,u)$, is constrained by the fluctuation-dissipation theorem (FDT) in the 
relativistic regime as  \cite{Ruggieri:2022kxv},
\begin{align}
\gamma(t,t')=\frac{{2\mathcal{D}}}{E(t')T}\frac{\langle h(t)h(t')\rangle}{\tau}.
\label{fdt}
\end{align}
Eq.~\eqref{Langevin_ph} differs from the standard Langevin equation in the scaling of the stochastic force term. In the present formulation, by contrast, the thermal noise is explicitly time correlated, so that the underlying dynamics are non-Markovian. This changes the scaling behavior of the stochastic term relative to the standard Markovian case. The time-discretized form of Eq.~\eqref{Langevin_ph}, 
used in the numerical implementation, is written as
\begin{align}
    \label{momnetum_HQ}p(t)= &p(t-\Delta t) - \Delta t \sum_{n=0}^{\textit{N-1}} \gamma(t,t'_n)p(t'_n)\Delta t  \nn & + \sqrt{\frac{2\mathcal{D}}{\tau}}h(t)\Delta t,
\end{align}
with notations, $t'_{0}= t_{0}$, $t'_{N-1}= t_{N} -\Delta t$ and $t'_{N}= t$. 
The numerical solution for the  HQ position evolution is given by
\begin{equation}\label{x_numericle}
    x(t_N)= x(t_{N-1}) + \left[\displaystyle\frac{p(t_{N-1})}{E(t)}\right]\Delta t.
\end{equation}
The momentum evolution of the HQs in the colored noise bath is determined by solving Eq.~\eqref{momnetum_HQ} simultaneously with Eq.~\eqref{SubA_x}. At each time  step, Eq.~\eqref{SubA_x} is calculated independently of Eq.~\eqref{momnetum_HQ}, subject to the initial condition $h(t=0) = 0$.

\subsubsection{Purely diffusive motion} 

For illustrative purposes, the one-dimensional (1-D) pure-diffusion condition is taken, although the actual numerical calculations are performed in three dimensions of the HQs momentum evolution. In this case, the drag coefficient in Eq.~\eqref{Langevin_p} is set to zero, and the initial condition is taken as \(p(0)=0\), so that the Langevin equation reduces to
%\begin{widetext}
\begin{align}
\frac{dp(t)}{dt} &= \xi(t).
\end{align}

%\eta(s)$ is a stationary Gaussian noise satisfying the contraction rule
%Contraction rule (used instead of writing a Dirac delta explicitly)
%\begin{align}
%&=\int_{0}^{t_1}\!\!\int_{0}^{t_2}
%\big\langle \eta(s_1)\,\eta(s_2)\big\rangle\, f(s_1,s_2)\,ds_1ds_2
%\\[-2pt]
%&= \tau \int_{0}^{\,\min(t_1,t_2)} f(s,s)\,ds .
%\end{align}
%The momentum at time $t$ is obtained by direct %integration of the Langevin equation,
% --- Step 1: Solve for p(t) ---

Together with the relation between \(\xi(t)\) and \(h(t)\) in Eq.~\eqref{h_rel}, the momentum at time \(t\) is obtained as
\begin{align}
p(t)
&= \int_{0}^{t}\xi(s)\,ds
= \sqrt{\frac{2\mathcal D}{\tau}}\int_{0}^{t} h(s)\,ds .
\end{align}
Using the definition of \(h(t)\) in Eq.~\eqref{h_evolution}, one finds
\begin{align}
p(t)
&= \sqrt{\frac{2\mathcal D}{\tau}}\;\frac{1}{\tau^{\nu}\Gamma(\nu)}
   \int_{0}^{t}\left[\int_{0}^{s}(s-u)^{\nu-1}\eta(u)\,du\right]ds.
\end{align}
Interchanging the order of integration gives
\begin{align}
p(t)
&= \sqrt{\frac{2\mathcal D}{\tau}}\;\frac{1}{\tau^{\nu}\Gamma(\nu)}
   \int_{0}^{t}\left[\int_{u}^{t}(s-u)^{\nu-1}ds\right]\eta(u)\,du .
\end{align}
The inner integral is elementary,
\begin{align}
\int_{u}^{t}(s-u)^{\nu-1}ds = \frac{(t-u)^{\nu}}{\nu} ,
\end{align}
so that
\begin{align}
p(t)
&= \sqrt{\frac{2\mathcal D}{\tau}}\;\frac{1}{\tau^{\nu}\nu\Gamma(\nu)}
   \int_{0}^{t}(t-u)^{\nu}\eta(u)\,du .
\end{align}
Using the identity \(\Gamma(\nu+1)=\nu\Gamma(\nu)\), this can be written in the compact form
\begin{align}\label{p_sol}
p(t)
&= \sqrt{\frac{2\mathcal D}{\tau}}\;\frac{1}{\tau^{\nu}\Gamma(\nu+1)}
   \int_{0}^{t}(t-u)^{\nu}\eta(u)\,du .
\end{align}

The unequal-time momentum covariance then follows from Eq.~\eqref{p_sol} together with the white-noise correlator used in Eq.~\eqref{corr_g} is, 
\begin{align}
\langle p(t_1)p(t_2)\rangle
=\frac{2\mathcal D}{\tau^{2\nu}\Gamma(\nu+1)^2}
   \int_{0}^{t_<}(t_1-s)^{\nu}(t_2-s)^{\nu}\,ds ,
\end{align}
where \(t_< \equiv \min(t_1,t_2)\). For \(t_1>t_2>0\), this reduces to
\begin{align}
\langle p(t_1)p(t_2)\rangle
&= \frac{2\mathcal D}{\tau^{2\nu}\Gamma(\nu+1)^2}
   \int_{0}^{t_2}(t_1-s)^{\nu}(t_2-s)^{\nu}\,ds .
\end{align}
The result for \(t_2>t_1\) follows by the interchange \(t_1\leftrightarrow t_2\).
Setting \(t_1=t_2=t\), one obtains the equal-time momentum variance,
\begin{align}
\langle p(t)^2\rangle
&= \frac{2\mathcal D}{\tau^{2\nu}\Gamma(\nu+1)^2}
   \int_{0}^{t}(t-s)^{2\nu}\,ds \nonumber\\
&= \frac{2\mathcal D}{\tau^{2\nu}\Gamma(\nu+1)^2}\,
   \frac{t^{2\nu+1}}{2\nu+1},
\end{align}
therefore,
\begin{align}\label{p_ana}
\langle p(t)^2\rangle
&= \frac{2\mathcal D}{\tau^{2\nu}}\;
   \frac{t^{2\nu+1}}{(2\nu+1)\Gamma(\nu+1)^2},
\qquad 0<\nu\leq 1.
\end{align}

\subsection{Numerical Validation}
\begin{figure}%[htp]
	\centering
     \includegraphics[scale = .39]{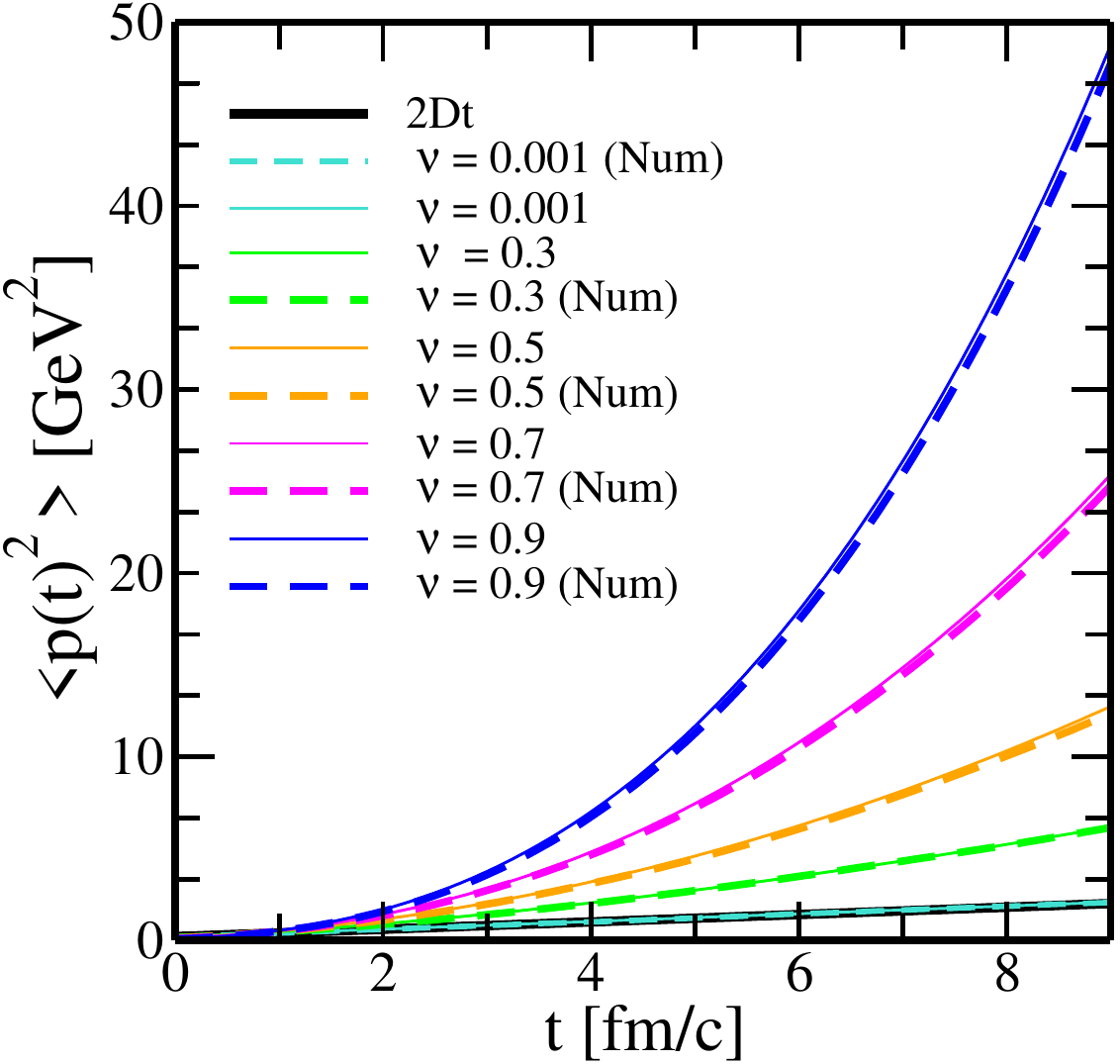}
        \caption{$\langle p^2(t) \rangle$ versus time for a 1-D, purely diffusive motion for different values of $\nu$ at $\tau = 1$ fm/c. Analytic solutions are represented by solid lines, and numerical (Num) solutions are represented by dashed lines.}
		\label{p_1D}
	\end{figure}
The numerical implementation of the memory kernel in Eq.~\eqref{momnetum_HQ}, constructed through the Caputo fractional derivative, is validated by direct comparison with the exact analytical expression for $\langle p^2(t) \rangle$ given in Eq.~\eqref{p_ana}. The comparison is performed in 1-D with purely diffusive dynamics, obtained by setting the drag coefficient, $\gamma = 0$ in Eq.~\eqref{momnetum_HQ}. In this paper, $\tau$ is the characteristic memory time,  which is associated with the noise correlation, and is taken to be 1~\text{fm}/c, the same value used in earlier work ~\cite{Ruggieri:2022kxv}. That work demonstrated that, even for a memory time of this magnitude, memory effects produce a measurable impact on the HQ observables,  the nuclear modification factor, $R_{AA}$. For the validation of the our numerical scheme, we consider, $\mathcal{D} = 0.1~\mathrm{GeV}^2/\mathrm{fm}$, the particle mass at $M = 1~\mathrm{GeV}$, and the memory index is varied over $\nu = 0.001,\, 0.3,\, 0.5,\, 0.7$, and $0.9$.

Figure~\ref{p_1D} presents a comparison between the analytical results 
(solid lines) given by Eq.~\eqref{p_ana} and the corresponding 
numerical results (dashed lines) obtained from Eq.~\eqref{momnetum_HQ} 
with $\gamma = 0$, for the time evolution of $\langle p^2(t)\rangle$  at several values of the memory parameter $\nu$. The black solid line, 
corresponding to $2\mathcal{D}t$, represents the Markovian limit of pure diffusion driven by white noise in the absence of memory. For 
very small values of $\nu$, both the numerical and analytical results reproduce $2\mathcal{D}t$, confirming that the white-noise limit is correctly recovered. Also, for all nonzero values of  $\nu$ considered, the numerical results are in close agreement with the corresponding analytical curves over the entire time interval.

This agreement validates the numerical implementation and confirms that the L1 scheme correctly incorporates the power-law time correlations of the thermal noise into the stochastic dynamics via the Caputo fractional derivative. The numerical scheme is therefore sufficiently reliable for the subsequent study of memory effects on the HQs momentum evolution in the QGP medium.

\section{Results}
\label{res}
In this section, we present the results for the normalized momentum correlator, average momentum squared, average squared displacement, average kinetic energy, and normalized central moments of the HQs. For illustrative purposes, a simplified initial condition is first adopted. Memory effects are further examined using a realistic initial condition for the HQs at $T = 250$~MeV and $T = 500$~MeV, while maintaining a constant 
diffusion coefficient $\mathcal{D}$.  The mass of the charm quarks ($M_c = 1.3 ~\mathrm{GeV}$) and the bottom quarks ($M_b = 4.5~\mathrm{GeV}$) are taken. All results are calculated by solving Eq.~\eqref{momnetum_HQ} and Eq.~\eqref{x_numericle}, consistently incorporating both the drag and diffusion transport coefficient to the three-dimensional numerical GLE.

\subsection{ Randomization of the heavy quark momentum}

\begin{figure}%[ht]
		\centering
        \includegraphics[height=7.5cm,width=8.cm]{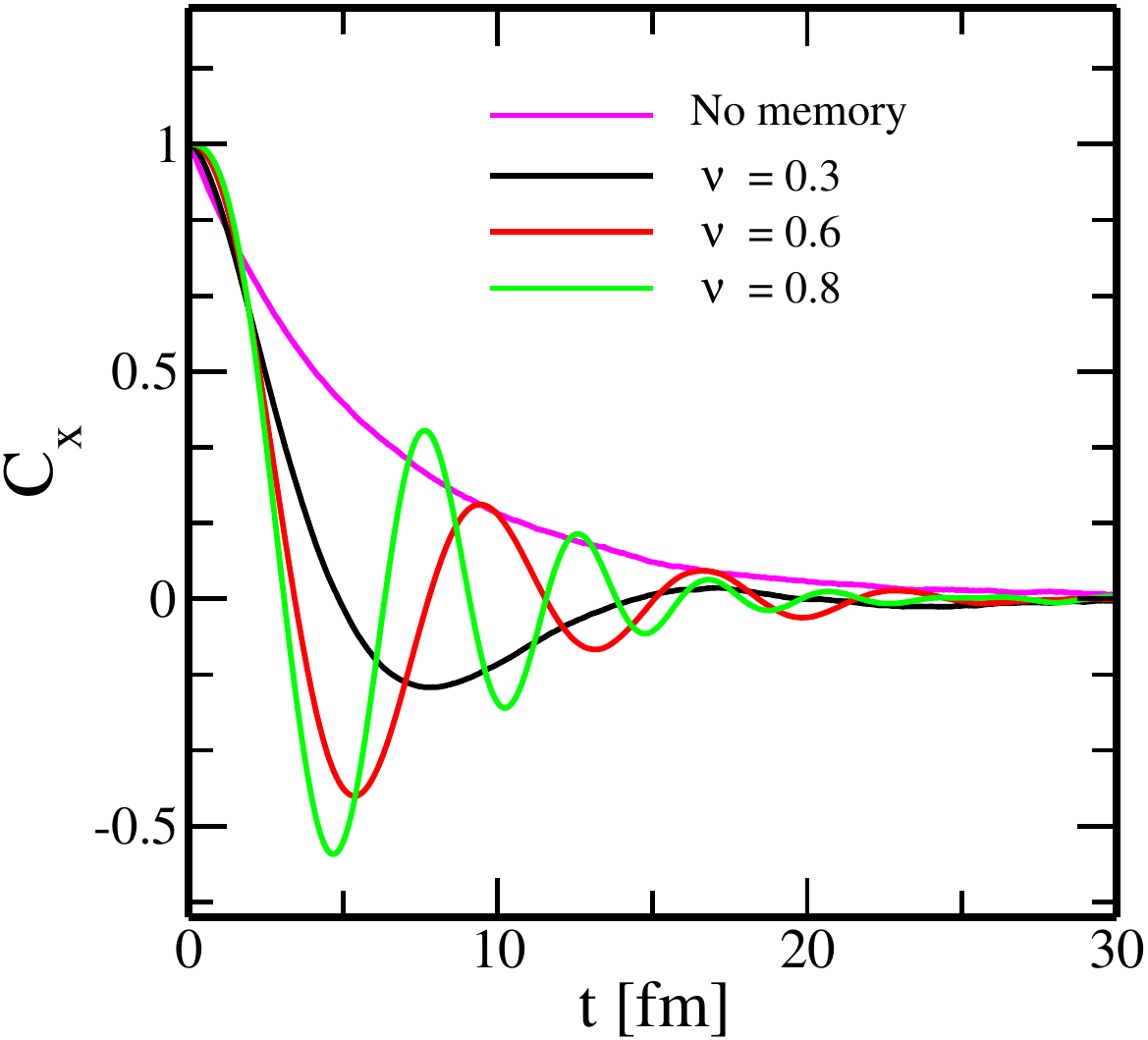}
        \caption{$C_x$ versus time of the HQs, considering the different values of $\nu$ at $\tau =1$ fm/c, $p_0$ = 1 GeV  and constant $\mathcal{D} = 0.2$ GeV$^2$/fm.}
		\label{correlation_p}
	\end{figure}

%%%%%%%%%%%%%%%%%%

The normalized momentum correlator is evaluated to examine the rate at which the HQs lose memory of their initial momentum while propagating through the QGP medium. It is defined as,
\begin{align}
C_x(t)=\frac{\langle p_x(t)\,p_x(0)\rangle}{\langle[p_x(0)]^2\rangle},
\end{align}
where $C_x(t)$ is the normalized momentum correlation function of the HQs momentum component $p_x$. The corresponding results for $C_y(t)$ and $C_z(t)$ in the $y$ and $z$ directions, respectively, are qualitatively similar. In Fig.~\ref{correlation_p}, $C_x(t)$ is shown for three representative values of the memory index, $\nu=0.3$, $0.6$, and $0.8$, at $T=500$~MeV and $\mathcal{D}=0.2$~GeV$^2$/fm.  When the time-correlated thermal noise is generated via the Caputo  process~\eqref{SubA_x}, the behavior of $C_x(t)$ depends sensitively 
on the memory parameter $\nu$. For reference, the Markovian case (magenta line) exhibits a smooth exponential decay toward zero, consistent with standard Langevin dynamics driven by uncorrelated 
thermal kicks~\cite{Moore:2004tg}. For weak memory ($\nu = 0.3$, black line), $C_x(t)$ remains close to this Markovian result, with only a marginal delay in the early-time decay. For stronger 
memory ($\nu = 0.6$ and $0.8$), the correlator develops a pronounced non-monotonic structure, including a negative lobe at intermediate times, before relaxing to $C_x(t) \approx 0$ at late times. This sign change reflects a transient reversal of momentum correlations 
induced by the continuation of the noise memory, an effect that is absent in the Markovian limit.
In the nonrelativistic limit, correlations of the type represented by $C_x$ have been studied broadly in stochastic systems with memory kernels \cite{WOS:000332486500022,WOS:000291660200006,SANDEV20113627}. 
This correlation thereby quantifies how medium interactions progressively suppress the initial momentum memory of the HQ.
Finally, we emphasize that $C_x(t)\to 0$ at late times for all $\nu$. This confirms that, once the FDT is implemented consistently, the HQs still approaches thermal equilibrium in the QGP; memory effects alter the path to equilibration by reshaping the early- and intermediate-time relaxation of momentum correlations.

%%%%%%%%%%%%%%%%%%%%%%%%%%%%%%%%%%%%%
\subsection{Effect of memory on  the average squared momentum and average squared displacement}

 \begin{figure*}[htp]
		\centering
        \includegraphics[scale = .37]{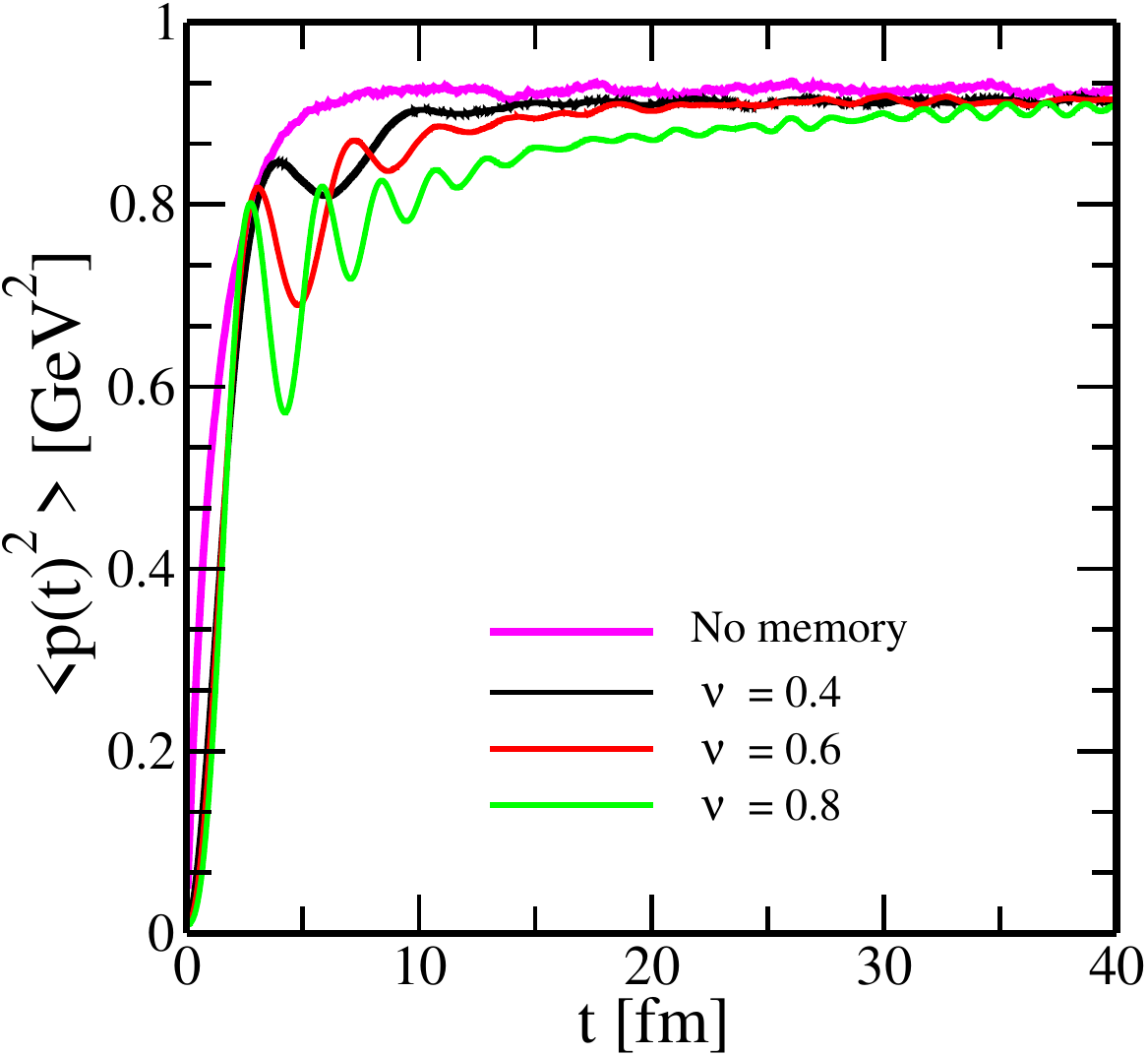}
         \hspace{10mm}
		\includegraphics[scale = .37]{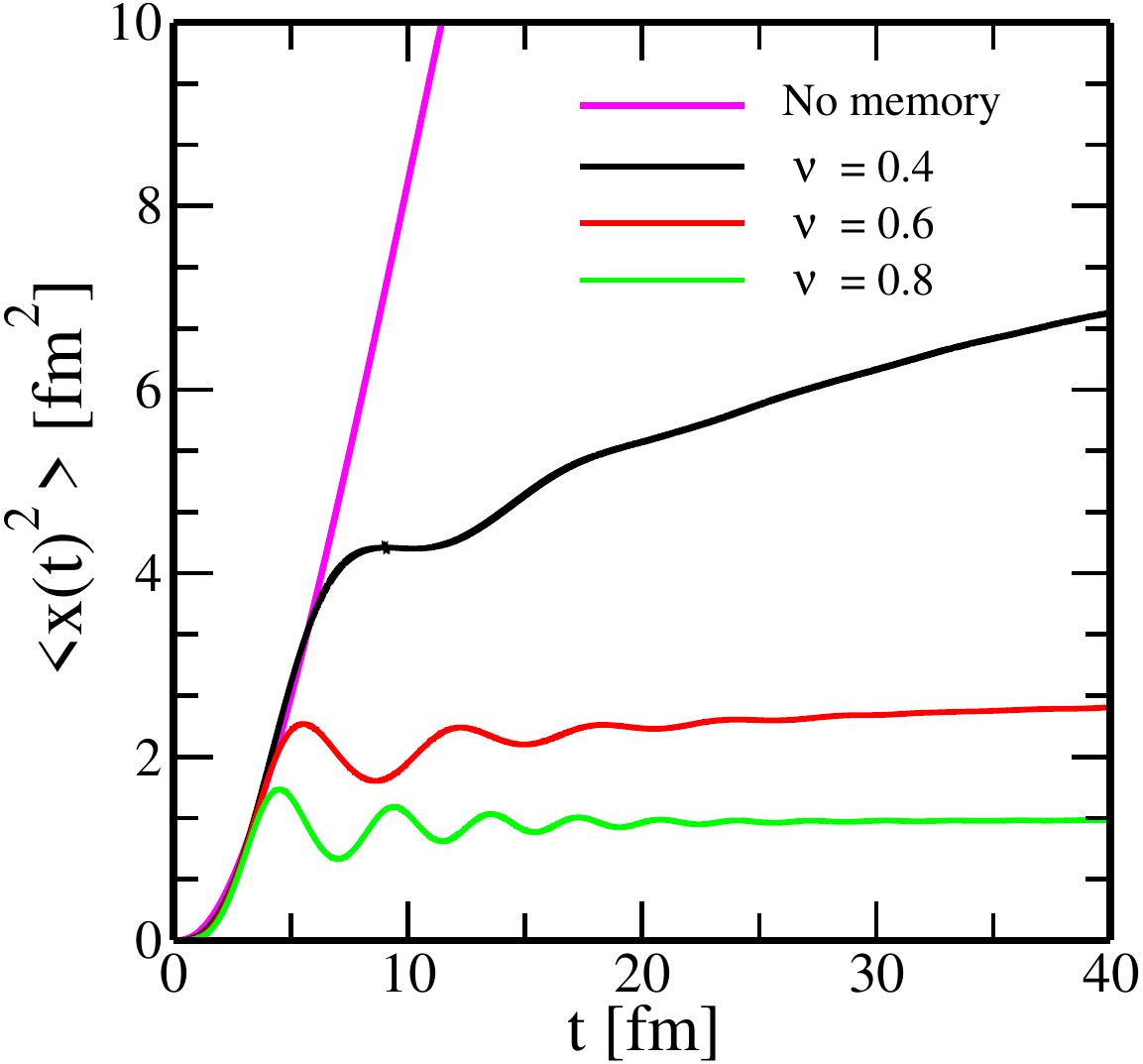}
        
		\caption{$\langle p^2(t) \rangle$ (left panel) and $\langle x^2(t) \rangle$ (right panel) versus time for the charm quark, for the various values of the $\nu$, $\mathcal{D} = 0.2$ {GeV}$^2$/fm, $\tau =1$ fm/c, and  $T = 250$ MeV. }
  
  \label{p_2D}
	\end{figure*}

\begin{figure*}[htp]
		\centering
        \includegraphics[scale = .37]{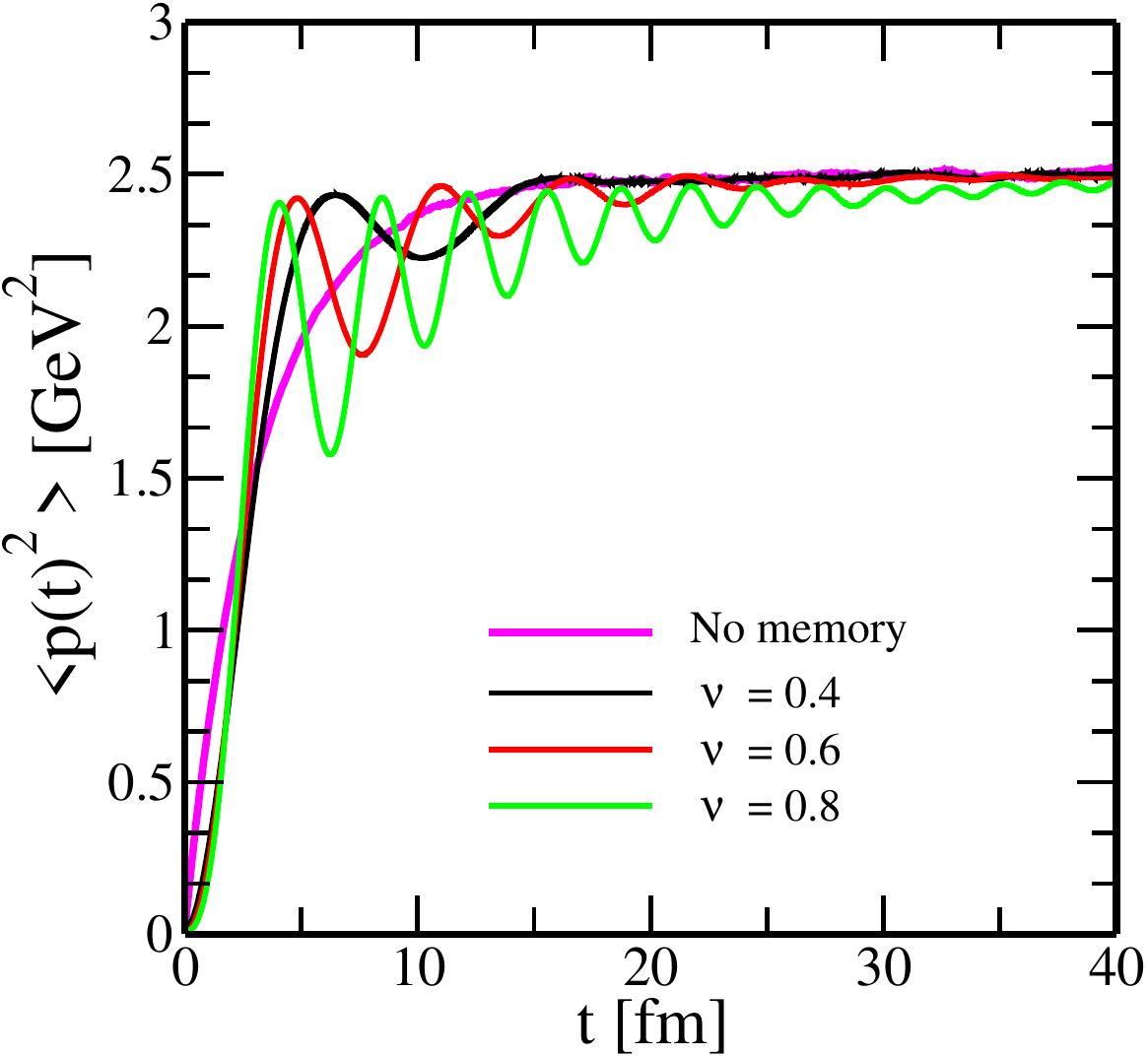}
         \hspace{10mm}
		\includegraphics[scale = .37]{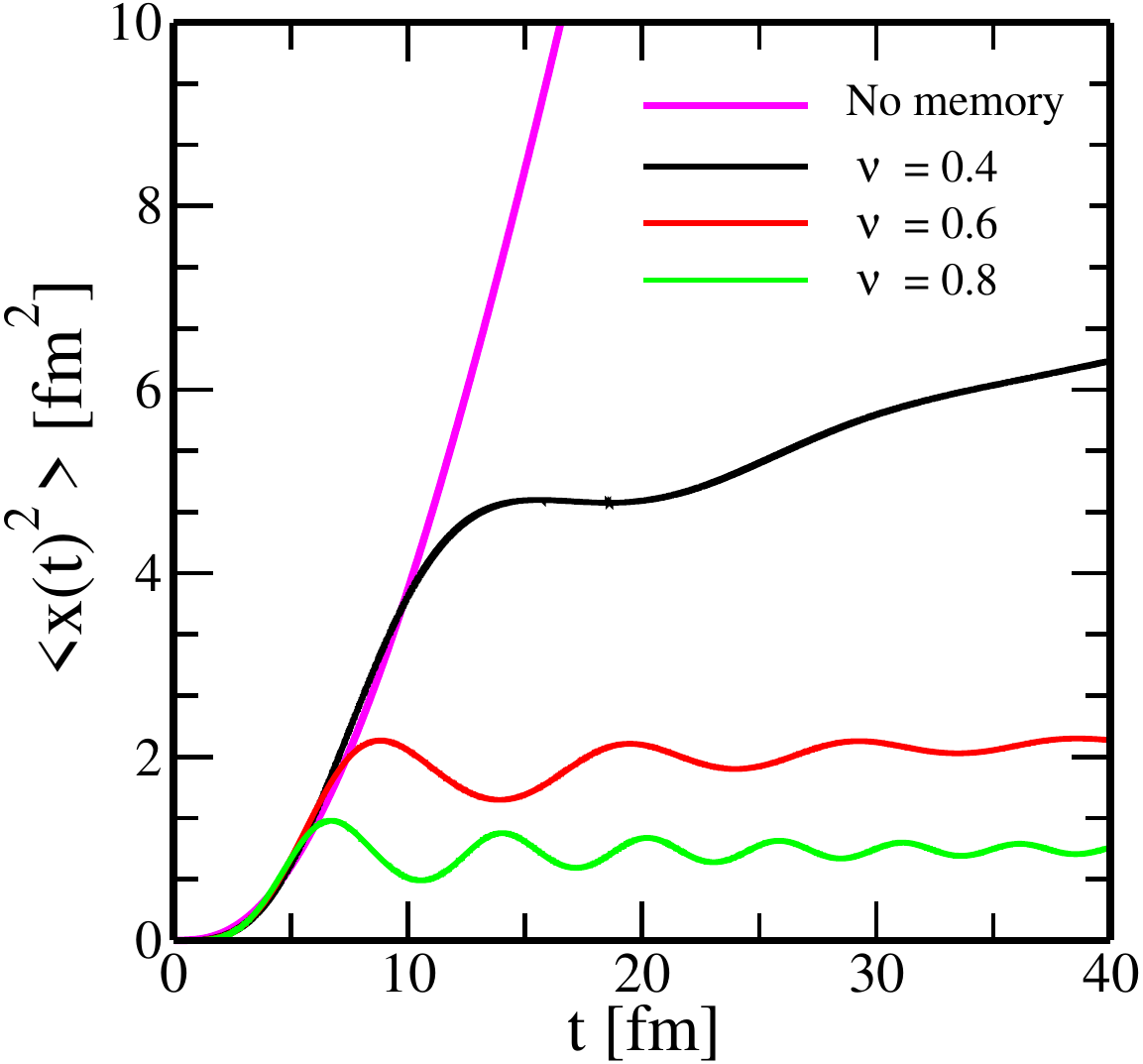}
        
		\caption{$\langle p^2(t) \rangle$ (left panel) and $\langle x^2(t) \rangle$ (right panel)  versus time for the bottom quark, for the various values of the $\nu$, $\mathcal{D} = 0.2$ {GeV}$^2$/fm, $\tau =1$ fm/c, and  $T = 250$ MeV. }
  
  \label{p_2D_B}
	\end{figure*}

For the results shown in Figs.~\ref{p_2D} and \ref{p_2D_B}, the HQs are initialized with transverse momentum $p_0=0.1$~GeV and propagated in a QGP medium characterized by $T=250$~MeV and $\mathcal{D}=0.2$~GeV$^2$/fm. The calculations are performed for three representative values of the memory index, $\nu=0.4$, $0.6$, and $0.8$, together with the no-memory case for comparison. The effect of memory on $\langle p^{2}(t)\rangle$ is shown in Fig.~\ref{p_2D} (left panel), where the time evolution of the transverse average squared momentum, defined as $\langle p^{2}(t)\rangle=\langle p_x^2(t)+p_y^2(t)\rangle$, is presented for the charm quark. In all cases, $\langle p^{2}(t)\rangle$ increases rapidly at early times. However, the inclusion of memory modifies the relaxation pattern qualitatively. Whereas the no-memory result approaches the asymptotic regime smoothly, the non-Markovian trajectories exhibit damped oscillations whose amplitude increases with $\nu$. Simultaneously, larger values of $\nu$ produce a stronger suppression of $\langle p^{2}(t)\rangle$ over the displayed time interval, indicating a slower approach to the asymptotic regime. This behavior reflects the increasing role of time correlations in thermal noise, which modify the HQ momentum evolution.

 The effect of memory on $\langle x^{2}(t)\rangle$ is calculated by Eq. \eqref{x_numericle}, which is shown in Fig.~\ref{p_2D} (right panel), where the time evolution of $\langle x^{2}(t)\rangle$ is shown for a charm quark using the same medium parameters and the same set of memory indices, $\nu=0.4,0.6,  0.8$, together with the no-memory case for comparison, with initial conditions $x(t_{0})=y(t_{0})=0$. A prominent dependence on $\nu$ is observed. For smaller $\nu$, the displacement grows more efficiently, whereas for larger $\nu$ the growth is substantially suppressed. The no-memory case exhibits the fastest growth and provides the corresponding Markovian reference~\cite{Moore:2004tg,Svetitsky:1987gq}. As $\nu$ decreases from $0.8$ to $0.4$, the late-time behavior of $\langle x^{2}(t)\rangle$ progressively approaches an approximately linear time dependence.

Fig. ~\ref{p_2D_B} shows the corresponding results for the bottom quark. The time evolution of $\langle p^{2}(t)\rangle$ exhibits a pronounced dependence on the memory index $\nu$. For $\nu = 0.4$, the approach to the asymptotic value is comparatively smooth, whereas larger values of $\nu$ produce stronger oscillations and a more markedly delayed approach to equilibrium. The qualitative pattern is therefore consistent with that observed for the charm quark: increasing $\nu$ enhances the non-Markovian character of the dynamics and causes the equilibration process to occur with more delay. The bottom-quark results additionally display a more persistent oscillatory structure over an extended time interval, indicating that memory effects remain visible over longer timescales than in the charm case.

An analogous trend is observed in the $\langle x^{2}(t)\rangle$ for the bottom quark using the same medium parameters and the same set of memory indices, $\nu=0.4,\,0.6,$ and $0.8$, together with the no-memory case for comparison, with initial conditions $x(t_{0})=y(t_{0})=0$. The $\langle x^{2}(t)\rangle$ is progressively suppressed as $\nu$ increases, and the overall magnitude of the displacement is smaller than in the charm case for all values of $\nu$ considered. For $\nu = 0.4$, $\langle x^{2}(t)\rangle$ grows steadily and approaches an approximately linear late-time behavior. For $\nu = 0.6$ and $\nu = 0.8$, the evolution exhibits a distinct early-time overshoot followed by damped oscillations, after which the growth rate is markedly reduced. Taken together, the behavior of $\langle p^{2}(t)\rangle$ and $\langle x^{2}(t)\rangle$ shows that memory effects modify the dynamical evolution of the HQs in the medium.

\subsection{Heavy quark thermalization from kinetic energy}
\begin{figure}%[ht]
		\centering
        \includegraphics[height=7.5cm,width=8.cm]{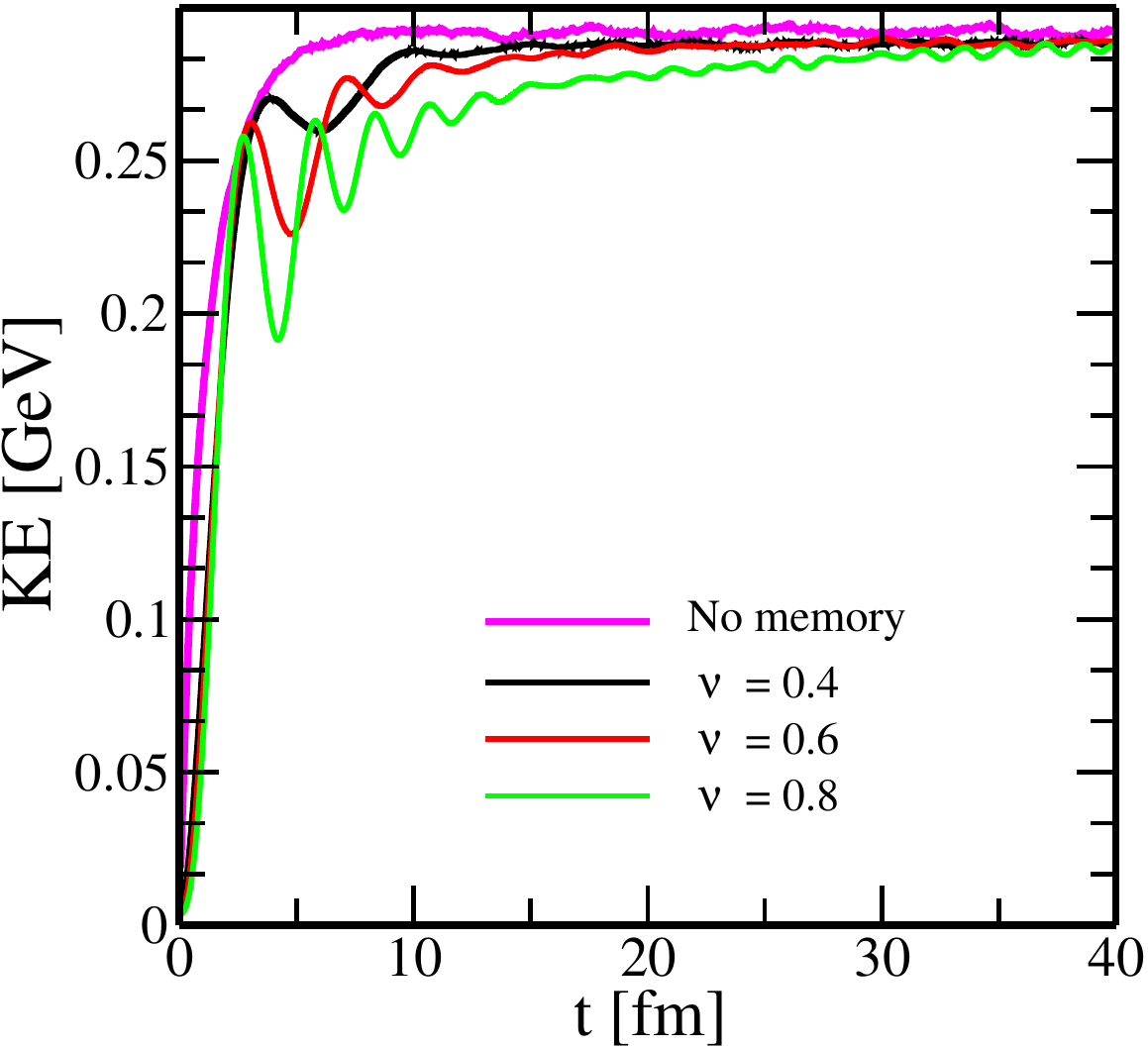}
        \caption{KE versus time of the HQs, considering the constant $\mathcal{D} = 0.2$ GeV$^2$/fm.}
		\label{kinetic}
	\end{figure}

To further characterize the approach to thermal equilibrium, the average kinetic energy (KE)  of the HQs is evaluated. This provides a direct measure of the extent to which the system approaches the thermal expectation set by the medium temperature. It is defined as, 
\begin{align}
    KE = \left\langle \sqrt{p^2+M_c^2}-M_c \right\rangle.
\end{align}
The calculations are performed at temperature $T = 250$~MeV, with charm-quark mass $M_c = 1.3$~GeV, $\tau =1$ fm/c, and  $\mathcal{D} = 0.2$~GeV$^2$/fm, consistent with parameter values commonly adopted in pQCD calculations. Figure~\ref{kinetic} displays the time evolution of the  KE of the charm quark for three representative values of the memory index, $\nu = 0.4$, $0.6$, and $0.8$, together with the memoryless reference case (magenta line). In all cases, $KE$ increases at early times and subsequently approaches a plateau, indicating the beginning of thermalization with the surrounding medium. The rate and character of this thermalization depend sensitively on $\nu$. For $\nu = 0.4$, the plateau is reached within the explored time interval and the evolution remains close to the memoryless result, indicating that weak memory produces only a marginal retardation of thermalization. As $\nu$ increases to $0.6$ and $0.8$, oscillatory structures develop in the $KE$ evolution and the approach to the plateau is progressively delayed. For $\nu = 0.8$, the asymptotic plateau is reached only at late times within the explored time interval, indicating that stronger memory 
substantially prolongs the thermalization timescale 
relative to the memoryless case.

These results demonstrate that increasing the memory parameter systematically retards the HQs thermalization, with the strength of this retardation growing monotonically with $\nu$. The bottom-quark results exhibit the same qualitative behavior and are not shown separately.
A quantitative estimate of the thermalization time under realistic initial conditions is presented in the following subsection. The subsequent analysis extends this study to phenomenologically relevant conditions.

%%%%%%%%%%%%%%%%%%%%%%%%%%%%%
\subsection{Memory effect for realistic initializations}

 \begin{figure*}[htp]
		\centering
        \includegraphics[scale = .37]{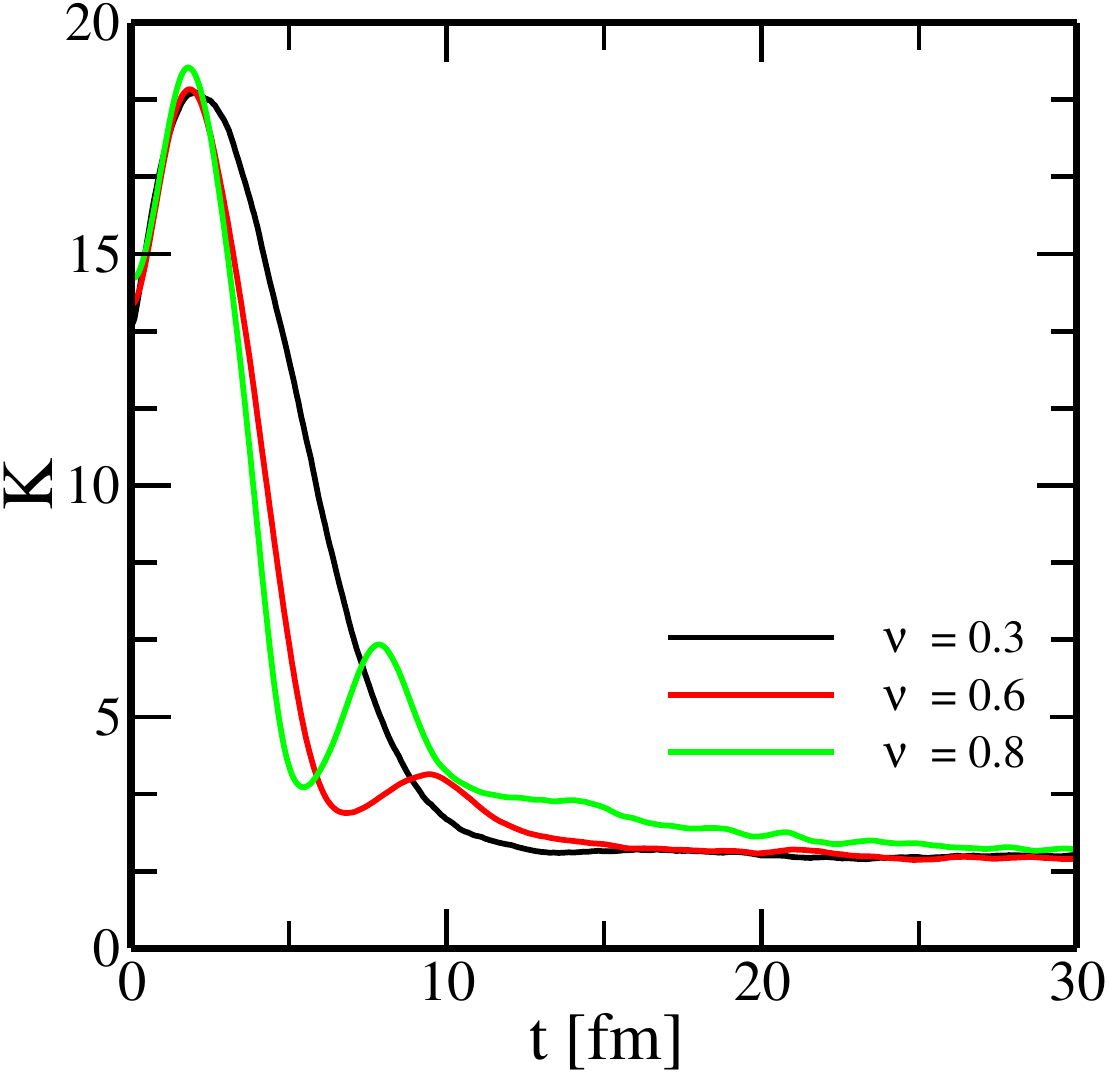}
         \hspace{10mm}
		\includegraphics[scale = .37]{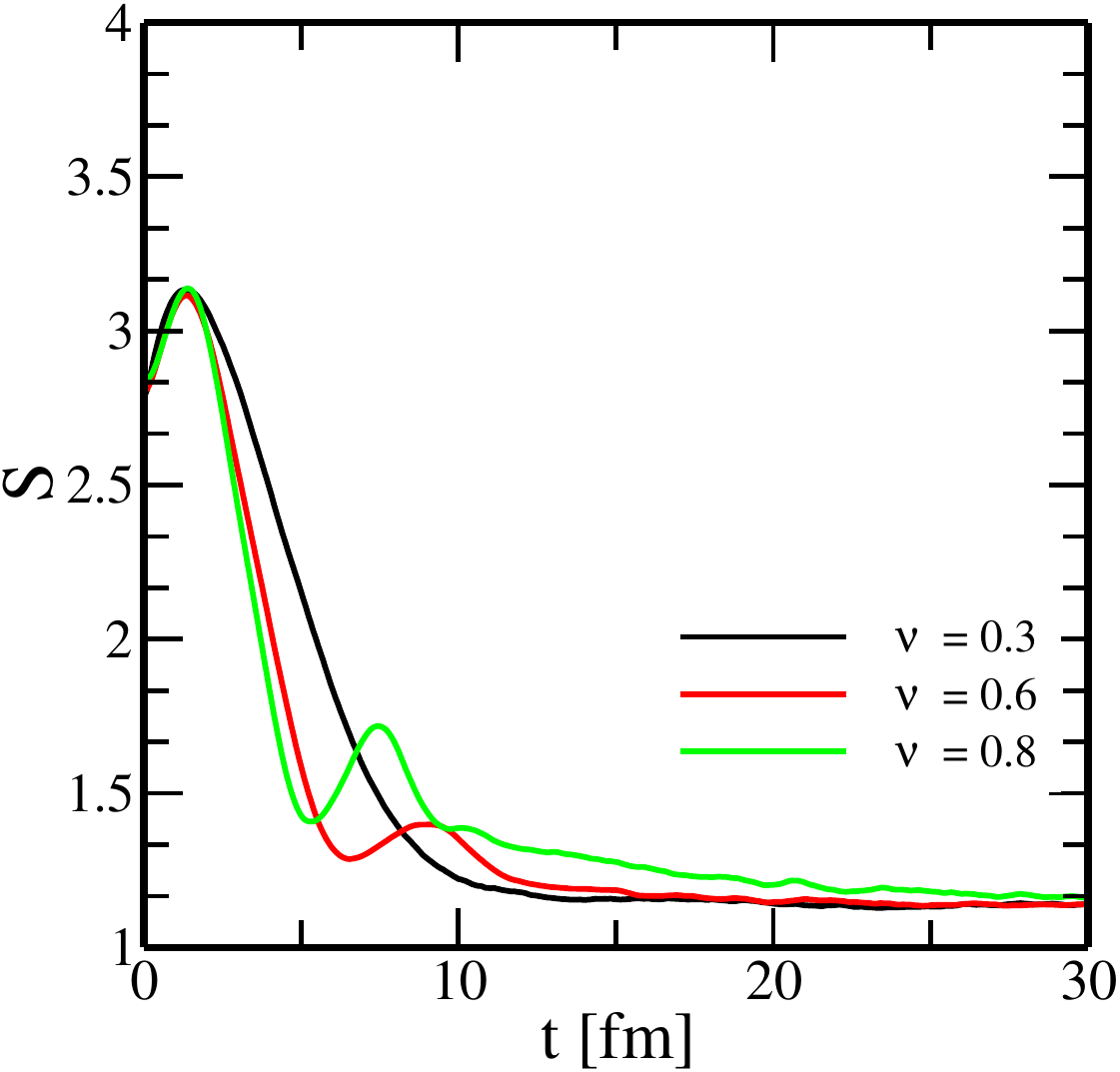}
        
		\caption{$K$ (left panel) and $S$ (right panel)  as functions of time for a charm quark at various values of $\nu$, with $\mathcal{D} = 0.2$ {GeV}$^2$/fm, $T = 500$ MeV and $\tau =1$ fm/c. }

		\label{cumulant}
	\end{figure*}

The initial condition for the charm quark transverse-momentum distribution is obtained 
from the fixed-order-plus-next-to-leading-logarithm (FONLL) form~\cite{Cacciari:2005rk,Cacciari:2012ny},
\begin{equation}
\frac{dN}{d^2p_T} = \frac{x_0}{\left(x_1 + p_T\right)^{x_2}},
\end{equation}
where $p_T$ denotes the HQs transverse momentum. The parameters are fixed to $x_0 = 6.36548 \times 10^8$, $x_1 = 9.0$, and $x_2 = 10.27890$. To characterize the effect of memory on the evolution of the realistic HQs transverse-momentum distribution, we study its higher normalized central moments. 
The non-Gaussian characteristics of the HQs transverse-momentum distribution are quantified through its higher normalized central moments. Specifically, the normalized third central moment characterizes the asymmetry of the distribution about its mean, whereas the  normalized fourth central moment describes the relative weight of the tails and how strongly the distribution is peaked around its mean value. For the HQs transverse-momentum distribution, these quantities are defined as
\begin{align}
 \quad S &=
\frac{\left\langle\left(p_T-\langle p_T\rangle\right)^3\right\rangle}
{\left\langle\left(p_T-\langle p_T\rangle\right)^2\right\rangle^{3/2}},
\label{skewness} \\[6pt]
 \quad K &=
\frac{\left\langle\left(p_T-\langle p_T\rangle\right)^4\right\rangle}
{\left\langle\left(p_T-\langle p_T\rangle\right)^2\right\rangle^{2}}
- 3,
\label{kurtosis}
\end{align}
where $p_T = \sqrt{p_x^2 + p_y^2}$ is the magnitude of the transverse momentum of the HQ, and $\langle p_T \rangle$ denotes an average over all particles at time $t$. $S$ and $K$ represent the normalized third central moment and the normalized fourth central moment, respectively. They provide information on the evolution of the HQs transverse-momentum distribution in the presence of memory.

Figure~\ref{cumulant} shows the time evolution of $S$ and $K$ for the HQs transverse-momentum distribution for three representative values of the memory parameter, $\nu=0.3$, $0.6$, and $0.8$. At $t=0$, both $S$ and $K$ assume large positive values, reflecting the pronounced high-$p_T$ tail of the FONLL initial spectrum and indicating that the distribution is strongly non-Gaussian at the beginning of the evolution. As the HQs propagate through the medium, both quantities decrease monotonically, indicating a gradual reduction of the initial asymmetry and tail weight through interactions with the thermal bath. The rate of this relaxation depends sensitively on $\nu$. For $\nu=0.3$, both $S$ and $K$ are suppressed comparatively rapidly, and the distribution evolves toward a more symmetric form within the time interval considered. By contrast, for $\nu=0.6$ and $0.8$, the decay becomes progressively slower, and the distribution retains its initial nonequilibrium character over longer timescales. This behaviour is consistent with the role of the memory parameter, according to which stronger time correlation in the noise delays the relaxation of the distribution toward equilibrium.
These results demonstrate that the higher normalized central moments  $S$ and $K$ are sensitive to the memory parameter $\nu$ and provide  a useful description of the influence of power-law time-correlated  thermal noise on the evolution of the HQs transverse-momentum 
distribution in the QGP medium.

%%%%%%%%%%%%%%%%%%

\section{Conclusion and outlook}
\label{con}

In this work, the effects of power-law time-correlated thermal noise on the 
HQs dynamics in the QGP have been studied within a  GLE framework. The time-correlated thermal noise is 
generated by a fractional differential equation formulated using the Caputo  fractional derivative of order $\nu$. The memory parameter $\nu$ controls both the strength and the power-law decay rate of the noise time correlations, with the Markovian white-noise limit recovered as $\nu \to 0$. The Caputo fractional derivative was discretized  using an L1 numerical scheme, and the resulting implementation was validated against the exact 
analytical solution for $\langle p^2(t)\rangle$ in a 
one-dimensional purely diffusive motion over the full range of  memory indices considered. The agreement obtained confirms the numerical reliability of the GLE for the subsequent analysis of the HQ dynamics in the QGP medium.

The HQ dynamics show that memory effects modify the  thermalization process in a systematic manner. The normalized momentum autocorrelation, $C_x(t)$, exhibits clear sensitivity to $\nu$: for weak memory, the behaviour remains close to the Markovian result, whereas for larger $\nu$ the correlator becomes non-monotonic and develops a negative lobe before relaxing 
to zero at late times, indicating that memory changes the thermalization  path by which initial momentum information is dissipated without preventing equilibration. Consistently, $\langle p^2(t)\rangle$  displays increasingly pronounced transient oscillations as $\nu$ 
increases, while $\langle x^2(t)\rangle$ is progressively 
suppressed, reflecting the retarded character of the non-Markovian  dynamics. The average kinetic energy, $KE$,  rises at early times and approaches a plateau in all cases, but stronger memory produces a more oscillatory relaxation 
pattern and a delayed approach to this asymptotic regime. These qualitative features are observed for both charm and bottom quarks. Taken together, these results show that time correlations in the thermal noise produce non-negligible modifications to the HQ dynamics.

To assess whether these effects continue under phenomenologically motivated initial conditions, the analysis was extended to the HQs initialized with the FONLL transverse-momentum distribution. The time evolution of $S$ and $K$ was then used to characterize memory-induced modifications of the shape of the $p_T$ spectrum. Both quantities decrease monotonically with time, reflecting a progressive reduction of the initial asymmetry and heavy-tail structure. This decrease becomes systematically slower as $\nu$ increases, indicating that the non-equilibrium character of the initial distribution is sustained over longer timescales. These results show that memory effects modify not only the thermalization process but also the shape of the HQs momentum distribution during its evolution in the medium.
The present results collectively establish that non-Markovian thermal noise leaves a characteristic and sizable impact on the HQ dynamics, manifested through delayed momentum correlation, oscillatory relaxation in $\langle p^2(t)\rangle$ and in the $KE$,  and a delayed relaxation of the transverse-momentum  distribution.

Several extensions of the present study are left for future work. A natural next step is the implementation of the present non-Markovian framework within an expanding QGP background with temperature-dependent transport coefficients, so that the memory effects identified here can be examined under more 
phenomenologically realistic conditions. The present framework considers a  fixed memory parameter $\nu$; extending the analysis to a time-dependent 
$\nu$, reflecting the evolving nature of the medium, would be a meaningful  generalization. A systematic study of how the non-Markovian dynamics 
modify the HQs spatial diffusion coefficient, $D_s$, and thermalization  time, as functions of both $\nu$ and $T$, would provide a more complete  quantitative characterization of memory-induced modifications to the HQ dynamics in the QGP medium.

\section{Acknowledgments} 

J. P. acknowledges Dr Santosh Kumar Das for valuable suggestions, Aditi Tomar for insightful discussions and encouragement, and Mohammad Yousuf Jamal for numerous informative contributions that have helped improve the content of this article.  This work was supported by the National Natural Science Foundation of China (NSFC) under Grant Nos. 12422508, 12375124, and the Science and Technology Commission of Shanghai Municipality under Grant No. 23JC1402700. Y.S. thanks the sponsorship from Yangyang Development Fund.

\section{Appendix}
\label{AP}
Consider a partition of the time interval $[0,T]$ given by ${t_n = \frac{n}{N}T : 0 \leq n \leq N}$.  For the scenario where $0<\nu\leq 1$, the two-step numerical scheme is employed \cite{Prakash:2024irm}. In this instance, the first derivative $u_1$ is approximated using a linear interpolation formula, resulting in a numerical scheme that solely relies on the values of $u$ at the preceding two time points $u_{n-1}$ and $u_{n-2}$, which is given as follows,
\begin{widetext}
\begin{align}\label{AP_sub}
\nonumber ^{C} D^\nu_{0+}u(t_n) &= \frac{1}{\Gamma({1-\nu})}\int_0^{t_n} \frac{u^{1}(s)}{(t_n-s)^{\nu}}ds \\
\nonumber&= \frac{1}{\Gamma({1-\nu})}\sum_{j=1}^n\int_{t_{j-1}}^{t_j} \frac{u^{1}(s)}{(t_n-s)^{\nu}}ds \\ 
\nonumber&\approx \frac{1}{\Gamma({1-\nu})}\sum_{j=1}^n\frac{u(t_j)-u(t_{j-1})}{t_j -t_{j-1}}\int_{t_{j-1}}^{t_j} \frac{1}{(t_n-s)^{\nu}}ds \\
\nonumber& =\frac{1}{\Gamma({2-\nu})} \sum_{j=1}^n \frac{(t_n-t_{j-1})^{1-\nu}- (t_n -t_j)^{1-\nu}}{t_j -t_{j-1}} (u(t_j)-u(t_{j-1}) )\\
& = \sum_{j=0}^{n-1} a_j (u(t_{n-j})-u(t_{n-j-1})),
\end{align}

\begin{align}\label{AP_super}
^{C} D^\nu_{0+}u(t_n)=\begin{cases}\displaystyle\frac{u(t_1)-u(t_0)}{\Delta t^{\nu}\Gamma(2-\nu)} \; &:\;n=1 \\
\displaystyle\frac{u(t_n)-u(t_{n-1})}{\Delta t^{\nu}\Gamma(2-\nu)} +\sum_{j=1}^{n-1} a_j(u(t_{n-j})-u(t_{n-j-1})) \; &:\;n \geq 2.
\end{cases}
\end{align}

\end{widetext}
where the coefficients {$a_j =\frac{((j+1)^{1-\nu}-j^{1-\nu})}{\Delta t^{\nu}\Gamma(2-\nu) }$} for $1 \leq j \leq n-1$ in the scheme are determined by the difference formula for the first 
derivative and are used to account for the fractional order.

\bibliography{ref1}

@article{li2013finite,
  title={The finite difference methods for fractional ordinary differential equations},
  author={Li, Changpin and Zeng, Fanhai},
  journal={Numerical Functional Analysis and Optimization},
  volume={34},
  number={2},
  pages={149--179},
  year={2013},
  publisher={Taylor \& Francis}
}

@article{miller1993introduction,
  title={An introduction to the fractional calculus and fractional
              differential equations},
  author={Miller, Kenneth S. and Ross, Bertram},
  journal={John Wiley \& Sons, Inc., New York},
  number={1219954},
  pages={xvi+366},
  year={1993},
  publisher={A Wiley-Interscience Publication}
}

@article{SANDEV20113627,
title = {Generalized Langevin equation with a three parameter Mittag-Leffler noise},
journal = {Physica A: Statistical Mechanics and its Applications},
volume = {390},
number = {21},
pages = {3627-3636},
year = {2011},
issn = {0378-4371},
doi = {https://doi.org/10.1016/j.physa.2011.05.039},
url = {https://www.sciencedirect.com/science/article/pii/S0378437111004456},
author = {Trifce Sandev and Živorad Tomovski and Johan L.A. Dubbeldam}}

@article{SandevMetzlerTomovski,
author = {Trifce Sandev and Ralf Metzler and Živorad Tomovski},
pages = {426--450},
volume = {15},
number = {3},
journal = {Fractional Calculus and Applied Analysis},
year = {2012},
lastchecked = {2024-03-27}
}

@article{Oliva:2024rex,
    author = "Oliva, Lucia and Parisi, Gabriele and Greco, Vincenzo and Ruggieri, Marco",
    title = "{Melting of cc{\textasciimacron} and bb{\textasciimacron} pairs in the pre-equilibrium stage of proton-nucleus collisions at the Large Hadron Collider}",
    eprint = "2412.07967",
    archivePrefix = "arXiv",
    primaryClass = "hep-ph",
    doi = "10.1103/nc3z-vns9",
    journal = "Phys. Rev. D",
    volume = "112",
    number = "1",
    pages = "014008",
    year = "2025"
}

@article{PhysRevC.49.1693,
  title = {Memory effects in relativistic heavy ion collisions},
  author = {Greiner, Carsten and Wagner, Klaus and Reinhard, Paul-Gerhard},
  journal = {Phys. Rev. C},
  volume = {49},
  issue = {3},
  pages = {1693--1701},
  numpages = {0},
  year = {1994},
  month = {Mar},
  publisher = {American Physical Society},
  doi = {10.1103/PhysRevC.49.1693},
  url = {https://link.aps.org/doi/10.1103/PhysRevC.49.1693}
}

@inproceedings{Mainardi2007SubdiffusionEO,
  title={Sub-diffusion equations of fractional order and their fundamental solutions},
  author={Francesco Mainardi and Antonio Mura and Gianni Pagnini and Rudolf Gorenflo and Enrico Clementel},
  year={2007},
  url={https://api.semanticscholar.org/CorpusID:55689962},
  doi={10.1007/978-1-4020-5678-9_3}
}

@article{PhysRevLett.93.180603,
  title = {Generalized Langevin Equation with Fractional Gaussian Noise: Subdiffusion within a Single Protein Molecule},
  author = {Kou, S. C. and Xie, X. Sunney},
  journal = {Phys. Rev. Lett.},
  volume = {93},
  issue = {18},
  pages = {180603},
  numpages = {4},
  year = {2004},
  month = {Oct},
  publisher = {American Physical Society},
  doi = {10.1103/PhysRevLett.93.180603},
  url = {https://link.aps.org/doi/10.1103/PhysRevLett.93.180603}
}

@article{Prakash:2024irm,
    author = "Prakash, Jai",
    title = "{Subdiffusion of heavy quarks in hot QCD matter by the fractional Langevin equation}",
    eprint = "2406.18714",
    archivePrefix = "arXiv",
    primaryClass = "hep-ph",
    doi = "10.1103/PhysRevC.110.044902",
    journal = "Phys. Rev. C",
    volume = "110",
    number = "4",
    pages = "044902",
    year = "2024"
}

@article{PhysRevE.107.064131,
  title = {Persistent nonequilibrium effects in generalized Langevin dynamics of nonrelativistic and relativistic particles},
  author = {Chen, Weiguo and Greiner, Carsten and Xu, Zhe},
  journal = {Phys. Rev. E},
  volume = {107},
  issue = {6},
  pages = {064131},
  numpages = {12},
  year = {2023},
  month = {Jun},
  publisher = {American Physical Society},
  doi = {10.1103/PhysRevE.107.064131},
  url = {https://link.aps.org/doi/10.1103/PhysRevE.107.064131}
}

@article{Jamal:2025gjy,
    author = "Jamal, Mohammad Yousuf and Li, Fu-Peng and Pang, Long-Gang and Qin, Guang-You",
    title = "{Melting of heavy quarkonia in QGP using deep neural networks}",
    eprint = "2509.14970",
    archivePrefix = "arXiv",
    primaryClass = "hep-ph",
    doi = "10.1103/z3kh-zr1t",
    journal = "Phys. Rev. C",
    volume = "113",
    number = "3",
    pages = "034915",
    year = "2026"
}

@article{Lim2009ModelingSD,
  title={Modeling single-file diffusion with step fractional Brownian motion and a generalized fractional Langevin equation},
  author={S. C. Lim and Lee Peng Teo},
  journal={Journal of Statistical Mechanics: Theory and Experiment},
  year={2009},
  volume={2009},
  pages={P08015},
  url={https://api.semanticscholar.org/CorpusID:118620117}
}

@article{STAR:2006vcp,
    author = "Adams, J. and others",
    collaboration = "STAR",
    title = "{Direct observation of dijets in central Au+Au collisions at s(NN)**(1/2) = 200-GeV}",
    eprint = "nucl-ex/0604018",
    archivePrefix = "arXiv",
    doi = "10.1103/PhysRevLett.97.162301",
    journal = "Phys. Rev. Lett.",
    volume = "97",
    pages = "162301",
    year = "2006"
}

@article{Adams:2005dq,
    author = "Adams, John and others",
    collaboration = "STAR",
    title = "{Experimental and theoretical challenges in the search for the quark gluon plasma: The STAR Collaboration's critical assessment of the evidence from RHIC collisions}",
    eprint = "nucl-ex/0501009",
    archivePrefix = "arXiv",
    doi = "10.1016/j.nuclphysa.2005.03.085",
    journal = "Nucl. Phys. A",
    volume = "757",
    pages = "102--183",
    year = "2005"
}

@article{PHENIX:2004vcz,
    author = "Adcox, K. and others",
    collaboration = "PHENIX",
    title = "{Formation of dense partonic matter in relativistic nucleus-nucleus collisions at RHIC: Experimental evaluation by the PHENIX collaboration}",
    eprint = "nucl-ex/0410003",
    archivePrefix = "arXiv",
    doi = "10.1016/j.nuclphysa.2005.03.086",
    journal = "Nucl. Phys. A",
    volume = "757",
    pages = "184--283",
    year = "2005"
}

@article{Svetitsky:1987gq,
    author = "Svetitsky, B.",
    title = "{Diffusion of charmed quarks in the quark-gluon plasma}",
    doi = "10.1103/PhysRevD.37.2484",
    journal = "Phys. Rev. D",
    volume = "37",
    pages = "2484--2491",
    year = "1988"
}

@article{BRAHMS:2005gow,
    author = "Arsene, I. and others",
    collaboration = "BRAHMS",
    title = "{Centrality dependent particle production at y=0 and y \textasciitilde{} 1 in Au + Au collisions at s(NN)**(1/2) = 200-GeV}",
    eprint = "nucl-ex/0503010",
    archivePrefix = "arXiv",
    doi = "10.1103/PhysRevC.72.014908",
    journal = "Phys. Rev. C",
    volume = "72",
    pages = "014908",
    year = "2005"
}

@article{vanHees:2004gq,
    author = "van Hees, Hendrik and Rapp, Ralf",
    title = "{Thermalization of heavy quarks in the quark-gluon plasma}",
    eprint = "nucl-th/0412015",
    archivePrefix = "arXiv",
    doi = "10.1103/PhysRevC.71.034907",
    journal = "Phys. Rev. C",
    volume = "71",
    pages = "034907",
    year = "2005"
}

@article{Moore:2004tg,
    author = "Moore, Guy D. and Teaney, Derek",
    title = "{How much do heavy quarks thermalize in a heavy ion collision?}",
    eprint = "hep-ph/0412346",
    archivePrefix = "arXiv",
    doi = "10.1103/PhysRevC.71.064904",
    journal = "Phys. Rev. C",
    volume = "71",
    pages = "064904",
    year = "2005"
}

@article{Das:2024vac,
    author = "Das, Santosh K. and Torres-Rincon, Juan M. and Rapp, Ralf",
    title = "{Charm and Bottom Hadrons in Hot Hadronic Matter}",
    eprint = "2406.13286",
    archivePrefix = "arXiv",
    primaryClass = "hep-ph",
    month = "6",
    year = "2024"
}

@article{Chandra:2024ron,
    author = "Chandra, Vinod and Das, Santosh K.",
    title = "{B-mesons as essential probes of hot QCD matter}",
    eprint = "2402.18870",
    archivePrefix = "arXiv",
    primaryClass = "hep-ph",
    doi = "10.1140/epjs/s11734-024-01123-4",
    journal = "Eur. Phys. J. ST",
    volume = "233",
    number = "2",
    pages = "429--438",
    year = "2024"
}

@article{Das:2022lqh,
    author = "Das, Santosh K. and others",
    title = "{Dynamics of Hot QCD Matter -- Current Status and Developments}",
    eprint = "2208.13440",
    archivePrefix = "arXiv",
    primaryClass = "nucl-th",
    doi = "10.1142/S0218301322500975",
    journal = "Int. J. Mod. Phys. E",
    volume = "31",
    pages = "12",
    year = "2022"
}

@article{Debnath:2023zet,
    author = "Debnath, Manas and Ghosh, Ritesh and Jamal, Mohammad Yousuf and Kurian, Manu and Prakash, Jai",
    title = "{Energy loss of a fast moving parton in Gribov-Zwanziger plasma}",
    eprint = "2311.16005",
    archivePrefix = "arXiv",
    primaryClass = "hep-ph",
    doi = "10.1103/PhysRevD.109.L011503",
    journal = "Phys. Rev. D",
    volume = "109",
    number = "1",
    pages = "L011503",
    year = "2024"
}

@article{Das:2024wht,
    author = "Das, Santosh K. and others",
    title = "{Dynamics of hot QCD matter 2024 {\textemdash} hard probes}",
    eprint = "2412.14026",
    archivePrefix = "arXiv",
    primaryClass = "nucl-ex",
    doi = "10.1142/S0218301325440033",
    journal = "Int. J. Mod. Phys. E",
    volume = "34",
    number = "07",
    pages = "2544003",
    year = "2025"
}

@article{Song:2015sfa,
    author = "Song, Taesoo and Berrehrah, Hamza and Cabrera, Daniel and Torres-Rincon, Juan M. and Tolos, Laura and Cassing, Wolfgang and Bratkovskaya, Elena",
    title = "{Tomography of the Quark-Gluon-Plasma by Charm Quarks}",
    eprint = "1503.03039",
    archivePrefix = "arXiv",
    primaryClass = "nucl-th",
    doi = "10.1103/PhysRevC.92.014910",
    journal = "Phys. Rev. C",
    volume = "92",
    number = "1",
    pages = "014910",
    year = "2015"
}

@article{Andronic:2015wma,
    author = "Andronic, A. and others",
    title = "{Heavy-flavour and quarkonium production in the LHC era: from proton\textendash{}proton to heavy-ion collisions}",
    eprint = "1506.03981",
    archivePrefix = "arXiv",
    primaryClass = "nucl-ex",
    doi = "10.1140/epjc/s10052-015-3819-5",
    journal = "Eur. Phys. J. C",
    volume = "76",
    number = "3",
    pages = "107",
    year = "2016"
}

@article{Dong:2019unq,
    author = "Dong, Xin and Greco, Vincenzo",
    title = "{Heavy quark production and properties of Quark\textendash{}Gluon Plasma}",
    doi = "10.1016/j.ppnp.2018.08.001",
    journal = "Prog. Part. Nucl. Phys.",
    volume = "104",
    pages = "97--141",
    year = "2019"
}

@article{Xu:2018gux,
    author = "Xu, Yingru and others",
    title = "{Resolving discrepancies in the estimation of heavy quark transport coefficients in relativistic heavy-ion collisions}",
    eprint = "1809.10734",
    archivePrefix = "arXiv",
    primaryClass = "nucl-th",
    doi = "10.1103/PhysRevC.99.014902",
    journal = "Phys. Rev. C",
    volume = "99",
    number = "1",
    pages = "014902",
    year = "2019"
}

@article{ALICE:2010khr,
    author = "Aamodt, K and others",
    collaboration = "ALICE",
    title = "{Charged-particle multiplicity density at mid-rapidity in central Pb-Pb collisions at $\sqrt{s_{NN}} = 2.76$ TeV}",
    eprint = "1011.3916",
    archivePrefix = "arXiv",
    primaryClass = "nucl-ex",
    reportNumber = "CERN-PH-EP-2010-060",
    doi = "10.1103/PhysRevLett.105.252301",
    journal = "Phys. Rev. Lett.",
    volume = "105",
    pages = "252301",
    year = "2010"
}

@article{Rapp:2018qla,
    author = "Beraudo, A. and others",
    editor = "Rapp, R. and Gossiaux, P. B. and Andronic, A. and Averbeck, R. and Masciocchi, S.",
    title = "{Extraction of Heavy-Flavor Transport Coefficients in QCD Matter}",
    eprint = "1803.03824",
    archivePrefix = "arXiv",
    primaryClass = "nucl-th",
    doi = "10.1016/j.nuclphysa.2018.09.002",
    journal = "Nucl. Phys. A",
    volume = "979",
    pages = "21--86",
    year = "2018"
}

@article{Prino:2016cni,
    author = "Prino, Francesco and Rapp, Ralf",
    title = "{Open Heavy Flavor in QCD Matter and in Nuclear Collisions}",
    eprint = "1603.00529",
    archivePrefix = "arXiv",
    primaryClass = "nucl-ex",
    doi = "10.1088/0954-3899/43/9/093002",
    journal = "J. Phys. G",
    volume = "43",
    number = "9",
    pages = "093002",
    year = "2016"
}

@article{Aarts:2016hap,
    author = "Aarts, G. and others",
    title = "{Heavy-flavor production and medium properties in high-energy nuclear collisions - What next?}",
    eprint = "1612.08032",
    archivePrefix = "arXiv",
    primaryClass = "nucl-th",
    doi = "10.1140/epja/i2017-12282-9",
    journal = "Eur. Phys. J. A",
    volume = "53",
    number = "5",
    pages = "93",
    year = "2017"
}

@article{Plumari:2017ntm,
    author = "Plumari, Salvatore and Minissale, Vincenzo and Das, Santosh K. and Coci, G. and Greco, V.",
    title = "{Charmed Hadrons from Coalescence plus Fragmentation in relativistic nucleus-nucleus collisions at RHIC and LHC}",
    eprint = "1712.00730",
    archivePrefix = "arXiv",
    primaryClass = "hep-ph",
    doi = "10.1140/epjc/s10052-018-5828-7",
    journal = "Eur. Phys. J. C",
    volume = "78",
    number = "4",
    pages = "348",
    year = "2018"
}

@article{vanHees:2007me,
    author = "van Hees, H. and Mannarelli, M. and Greco, V. and Rapp, R.",
    title = "{Nonperturbative heavy-quark diffusion in the quark-gluon plasma}",
    eprint = "0709.2884",
    archivePrefix = "arXiv",
    primaryClass = "hep-ph",
    doi = "10.1103/PhysRevLett.100.192301",
    journal = "Phys. Rev. Lett.",
    volume = "100",
    pages = "192301",
    year = "2008"
}

@inproceedings{Rapp:2009my,
    author = "Rapp, Ralf and van Hees, Hendrik",
    title = "{Heavy Quarks in the Quark-Gluon Plasma}",
    eprint = "0903.1096",
    archivePrefix = "arXiv",
    primaryClass = "hep-ph",
    doi = "10.1142/9789814293297_0003",
    pages = "111--206",
    year = "2010"
}

@article{Singh:2023smw,
    author = "Singh, Mayank and Kurian, Manu and Jeon, Sangyong and Gale, Charles",
    title = "{Open charm phenomenology with a multistage approach to relativistic heavy-ion collisions}",
    eprint = "2306.09514",
    archivePrefix = "arXiv",
    primaryClass = "nucl-th",
    doi = "10.1103/PhysRevC.108.054901",
    journal = "Phys. Rev. C",
    volume = "108",
    number = "5",
    pages = "054901",
    year = "2023"
}

@article{Kurian:2020orp,
    author = "Kurian, Manu and Singh, Mayank and Chandra, Vinod and Jeon, Sangyong and Gale, Charles",
    title = "{Charm quark dynamics in quark-gluon plasma with 3 + 1D viscous hydrodynamics}",
    eprint = "2007.07705",
    archivePrefix = "arXiv",
    primaryClass = "hep-ph",
    doi = "10.1103/PhysRevC.102.044907",
    journal = "Phys. Rev. C",
    volume = "102",
    number = "4",
    pages = "044907",
    year = "2020"
}

@article{Ruggieri:2022kxv,
    author = "Ruggieri, Marco and Pooja and Prakash, Jai and Das, Santosh K.",
    title = "{Memory effects on energy loss and diffusion of heavy quarks in the quark-gluon plasma}",
    eprint = "2203.06712",
    archivePrefix = "arXiv",
    primaryClass = "hep-ph",
    doi = "10.1103/PhysRevD.106.034032",
    journal = "Phys. Rev. D",
    volume = "106",
    number = "3",
    pages = "034032",
    year = "2022"
}

@article{Pooja:2023gqt,
    author = "Pooja and Das, Santosh K. and Greco, Vincenzo and Ruggieri, Marco",
    title = "{Thermalization and isotropization of heavy quarks in a non-Markovian medium in high-energy nuclear collisions}",
    eprint = "2306.13749",
    archivePrefix = "arXiv",
    primaryClass = "hep-ph",
    doi = "10.1103/PhysRevD.108.054026",
    journal = "Phys. Rev. D",
    volume = "108",
    number = "5",
    pages = "054026",
    year = "2023"
}

@article{PhysRevLett.98.022301,
  title = {Dilepton Yields from Brown-Rho Scaled Vector Mesons Including Memory Effects},
  author = {Schenke, Bj\"orn and Greiner, Carsten},
  journal = {Phys. Rev. Lett.},
  volume = {98},
  issue = {2},
  pages = {022301},
  numpages = {4},
  year = {2007},
  month = {Jan},
  publisher = {American Physical Society},
  doi = {10.1103/PhysRevLett.98.022301},
  url = {https://link.aps.org/doi/10.1103/PhysRevLett.98.022301}
}

@article{PhysRevC.85.054906,
  title = {Relativistic theory of hydrodynamic fluctuations with applications to heavy-ion collisions},
  author = {Kapusta, J. I. and M\"uller, B. and Stephanov, M.},
  journal = {Phys. Rev. C},
  volume = {85},
  issue = {5},
  pages = {054906},
  numpages = {17},
  year = {2012},
  month = {May},
  publisher = {American Physical Society},
  doi = {10.1103/PhysRevC.85.054906},
  url = {https://link.aps.org/doi/10.1103/PhysRevC.85.054906}
}

@article{Ruggieri:2019zos,
    author = "Ruggieri, Marco and Frasca, Marco and Das, Santosh Kumar",
    title = "{Classical model for diffusion and thermalization of heavy quarks in a hot medium: memory and out-of-equilibrium effects}",
    eprint = "1903.11302",
    archivePrefix = "arXiv",
    primaryClass = "nucl-th",
    doi = "10.1088/1674-1137/43/9/094105",
    journal = "Chin. Phys. C",
    volume = "43",
    number = "9",
    pages = "094105",
    year = "2019"
}

@article{Schuller:2019ega,
    author = {Sch\"uller, B. and Meistrenko, A. and Van Hees, H. and Xu, Z. and Greiner, C.},
    title = "{Kramers\textquoteright{} escape rate problem within a non-Markovian description}",
    eprint = "1905.09652",
    archivePrefix = "arXiv",
    primaryClass = "cond-mat.stat-mech",
    doi = "10.1016/j.aop.2019.168045",
    journal = "Annals Phys.",
    volume = "412",
    pages = "168045",
    year = "2020"
}

@article{Murase:2013tma,
    author = "Murase, Koichi and Hirano, Tetsufumi",
    title = "{Relativistic fluctuating hydrodynamics with memory functions and colored noises}",
    eprint = "1304.3243",
    archivePrefix = "arXiv",
    primaryClass = "nucl-th",
    month = "4",
    year = "2013"
}

@article{Hammelmann:2018ath,
    author = "Hammelmann, Jan and Torres-Rincon, Juan M. and Rose, Jean-Bernard and Greif, Moritz and Elfner, Hannah",
    title = "{Electrical conductivity and relaxation via colored noise in a hadronic gas}",
    eprint = "1810.12527",
    archivePrefix = "arXiv",
    primaryClass = "hep-ph",
    doi = "10.1103/PhysRevD.99.076015",
    journal = "Phys. Rev. D",
    volume = "99",
    number = "7",
    pages = "076015",
    year = "2019"
}

@article{Gegechkori:2008ppw,
    author = "Gegechkori, A. E. and Anischenko, Yu. A. and Nadtochy, P. N. and Adeev, G. D.",
    title = "{Impact of non-Markovian effects on the fission rate and time}",
    doi = "10.1134/S1063778808120028",
    journal = "Phys. Atom. Nucl.",
    volume = "71",
    number = "12",
    pages = "2007--2017",
    year = "2008"
}

@article{Ivanyuk:2021tzw,
    author = "Ivanyuk, F. A. and Radionov, S. V. and Ishizuka, C. and Chiba, S.",
    title = "{Memory effects in Langevin approach to the nuclear fission process}",
    eprint = "2103.14145",
    archivePrefix = "arXiv",
    primaryClass = "nucl-th",
    month = "3",
    year = "2021"
}

@article{Das:2013kea,
    author = "Das, Santosh K. and Scardina, Francesco and Plumari, Salvatore and Greco, Vincenzo",
    title = "{Heavy-flavor in-medium momentum evolution: Langevin versus Boltzmann approach}",
    eprint = "1312.6857",
    archivePrefix = "arXiv",
    primaryClass = "nucl-th",
    doi = "10.1103/PhysRevC.90.044901",
    journal = "Phys. Rev. C",
    volume = "90",
    pages = "044901",
    year = "2014"
}

@article{Kumar:2021goi,
    author = "Kumar, Avdhesh and Kurian, Manu and Das, Santosh K. and Chandra, Vinod",
    title = "{Drag of heavy quarks in an anisotropic QCD medium beyond the static limit}",
    eprint = "2111.07563",
    archivePrefix = "arXiv",
    primaryClass = "hep-ph",
    doi = "10.1103/PhysRevC.105.054903",
    journal = "Phys. Rev. C",
    volume = "105",
    number = "5",
    pages = "054903",
    year = "2022"
}

@article{Prakash:2021lwt,
    author = "Prakash, Jai and Kurian, Manu and Das, Santosh K. and Chandra, Vinod",
    title = "{Heavy quark transport in an anisotropic hot QCD medium: Collisional and Radiative processes}",
    eprint = "2102.07082",
    archivePrefix = "arXiv",
    primaryClass = "hep-ph",
    doi = "10.1103/PhysRevD.103.094009",
    journal = "Phys. Rev. D",
    volume = "103",
    number = "9",
    pages = "094009",
    year = "2021"
}

@article{Prakash:2023wbs,
    author = "Prakash, Jai and Chandra, Vinod and Das, Santosh K.",
    title = "{Heavy quark radiation in an anisotropic hot QCD medium}",
    eprint = "2306.07966",
    archivePrefix = "arXiv",
    primaryClass = "hep-ph",
    doi = "10.1103/PhysRevD.108.096016",
    journal = "Phys. Rev. D",
    volume = "108",
    number = "9",
    pages = "096016",
    year = "2023"
}

@article{Jamal:2023ncn,
  author = {Jamal, M. Y. and Prakash, J. and Nilima, I. and Bandyopadhyay, A.},
  title = {Energy loss of heavy quarks in the presence of magnetic field},
  archivePrefix = {arXiv},
  eprint = {2304.09851},
  primaryClass = {hep-ph},
  year = {2023},
}

@article{Mazumder:2011nj,
    author = "Mazumder, Surasree and Bhattacharyya, Trambak and Alam, Jan-e and Das, Santosh K",
    title = "{Momentum dependence of drag coefficients and heavy flavour suppression in quark gluon plasma}",
    eprint = "1106.2615",
    archivePrefix = "arXiv",
    primaryClass = "nucl-th",
    doi = "10.1103/PhysRevC.84.044901",
    journal = "Phys. Rev. C",
    volume = "84",
    pages = "044901",
    year = "2011"
}

@article{scardina2017estimating,
  title={Estimating the charm quark diffusion coefficient and thermalization time from D meson spectra at energies available at the BNL Relativistic Heavy Ion Collider and the CERN Large Hadron Collider},
  author={Scardina, Francesco and Das, Santosh K and Minissale, Vincenzo and Plumari, Salvatore and Greco, Vincenzo},
  journal={Physical Review C},
  volume={96},
  number={4},
  pages={044905},
  year={2017},
  publisher={APS}
}

@article{Cao:2018ews,
    author = "Cao, Shanshan and others",
    title = "{Toward the determination of heavy-quark transport coefficients in quark-gluon plasma}",
    eprint = "1809.07894",
    archivePrefix = "arXiv",
    primaryClass = "nucl-th",
    doi = "10.1103/PhysRevC.99.054907",
    journal = "Phys. Rev. C",
    volume = "99",
    number = "5",
    pages = "054907",
    year = "2019"
}

@article{Jamal:2021btg,
    author = "Jamal, Mohammad Yousuf and Mohanty, Bedangadas",
    title = "{Passage of heavy quarks through the fluctuating hot QCD medium}",
    eprint = "2101.00164",
    archivePrefix = "arXiv",
    primaryClass = "nucl-th",
    doi = "10.1140/epjc/s10052-021-09418-9",
    journal = "Eur. Phys. J. C",
    volume = "81",
    number = "7",
    pages = "616",
    year = "2021"
}

@article{ZACCONE2024116483,
title = {Theory of heavy-quarks contribution to the quark-gluon plasma viscosity},
journal = {Nuclear Physics B},
volume = {1000},
pages = {116483},
year = {2024},
issn = {0550-3213},
doi = {https://doi.org/10.1016/j.nuclphysb.2024.116483},
url = {https://www.sciencedirect.com/science/article/pii/S055032132400049X},
author = {Alessio Zaccone},

}

@article{Das:2025yxy,
    author = "Das, Santosh K. and Soloveva, Olga and Song, Taesoo and Bratkovskaya, Elena",
    title = "{Influence of electromagnetic fields on the generation of the directed and elliptic flows of heavy quarks in relativistic heavy-ion collisions}",
    eprint = "2507.22620",
    archivePrefix = "arXiv",
    primaryClass = "nucl-th",
    doi = "10.1103/yppc-ts4n",
    journal = "Phys. Rev. C",
    volume = "112",
    number = "6",
    pages = "064901",
    year = "2025"
}

@article{Minissale:2023dct,
    author = "Minissale, Vincenzo and Plumari, Salvatore and Sun, Yifeng and Greco, Vincenzo",
    title = "{Multi-charmed and singled charmed hadrons from coalescence: yields and ratios in different collision systems at LHC}",
    eprint = "2305.03687",
    archivePrefix = "arXiv",
    primaryClass = "hep-ph",
    doi = "10.1140/epjc/s10052-024-12571-6",
    journal = "Eur. Phys. J. C",
    volume = "84",
    number = "3",
    pages = "228",
    year = "2024"
}

@article{Dey:2025kqx,
    author = "Dey, Debarshi and Bandyopadhyay, Aritra and Das, Santosh K. and Dash, Sadhana and Chandra, Vinod and Nandi, Basanta K.",
    title = "{Nonperturbative heavy quark diffusion coefficients in a weakly magnetized thermal QCD medium}",
    eprint = "2504.02284",
    archivePrefix = "arXiv",
    primaryClass = "hep-ph",
    doi = "10.1103/62yt-5r65",
    journal = "Phys. Rev. D",
    volume = "112",
    number = "1",
    pages = "016011",
    year = "2025"
}

@article{Sambataro:2025pop,
    author = "Sambataro, Maria Lucia and Plumari, Salvatore and Das, Santosh K. and Greco, Vincenzo",
    title = "{Probing the QGP through $p_T$-differential radial flow of heavy quarks}",
    eprint = "2510.19448",
    archivePrefix = "arXiv",
    primaryClass = "hep-ph",
    month = "10",
    year = "2025"
}

@article{PhysRevLett.130.231902,
  title = {Heavy Quark Diffusion from $2+1$ Flavor Lattice QCD with 320 MeV Pion Mass},
  author = {Altenkort, Luis and Kaczmarek, Olaf and Larsen, Rasmus and Mukherjee, Swagato and Petreczky, Peter and Shu, Hai-Tao and Stendebach, Simon},
  collaboration = {HotQCD Collaboration},
  journal = {Phys. Rev. Lett.},
  volume = {130},
  issue = {23},
  pages = {231902},
  numpages = {6},
  year = {2023},
  month = {Jun},
  publisher = {American Physical Society},
  doi = {10.1103/PhysRevLett.130.231902},
  url = {https://link.aps.org/doi/10.1103/PhysRevLett.130.231902}
}

@article{Cao:2016gvr,
    author = "Cao, Shanshan and Luo, Tan and Qin, Guang-You and Wang, Xin-Nian",
    title = "{Linearized Boltzmann transport model for jet propagation in the quark-gluon plasma: Heavy quark evolution}",
    eprint = "1605.06447",
    archivePrefix = "arXiv",
    primaryClass = "nucl-th",
    doi = "10.1103/PhysRevC.94.014909",
    journal = "Phys. Rev. C",
    volume = "94",
    number = "1",
    pages = "014909",
    year = "2016"
}

@article{Gossiaux:2008jv,
    author = "Gossiaux, P. B. and Aichelin, J.",
    editor = "Alam, Jan-e and Chattopadhyay, Subhasis and Nayak, Tapan and Sinha, Bikash and Viyogi, Yogendra P.",
    title = "{Towards an understanding of the RHIC single electron data}",
    eprint = "0802.2525",
    archivePrefix = "arXiv",
    primaryClass = "hep-ph",
    doi = "10.1103/PhysRevC.78.014904",
    journal = "Phys. Rev. C",
    volume = "78",
    pages = "014904",
    year = "2008"
}

@article{Cao:2015hia,
    author = "Cao, Shanshan and Qin, Guang-You and Bass, Steffen A.",
    title = "{Energy loss, hadronization and hadronic interactions of heavy flavors in relativistic heavy-ion collisions}",
    eprint = "1505.01413",
    archivePrefix = "arXiv",
    primaryClass = "nucl-th",
    doi = "10.1103/PhysRevC.92.024907",
    journal = "Phys. Rev. C",
    volume = "92",
    number = "2",
    pages = "024907",
    year = "2015"
}

@article{Cacciari:2005rk,
    author = "Cacciari, Matteo and Nason, Paolo and Vogt, Ramona",
    title = "{QCD predictions for charm and bottom production at RHIC}",
    eprint = "hep-ph/0502203",
    archivePrefix = "arXiv",
    reportNumber = "LPTHE-05-03, BIOCOCCA-FT-05-4, LBNL-57063, BICOCCA-FT-05-4",
    doi = "10.1103/PhysRevLett.95.122001",
    journal = "Phys. Rev. Lett.",
    volume = "95",
    pages = "122001",
    year = "2005"
}

@article{Cacciari:2012ny,
    author = "Cacciari, Matteo and Frixione, Stefano and Houdeau, Nicolas and Mangano, Michelangelo L. and Nason, Paolo and Ridolfi, Giovanni",
    title = "{Theoretical predictions for charm and bottom production at the LHC}",
    eprint = "1205.6344",
    archivePrefix = "arXiv",
    primaryClass = "hep-ph",
    reportNumber = "CERN-PH-TH-2011-227",
    doi = "10.1007/JHEP10(2012)137",
    journal = "JHEP",
    volume = "10",
    pages = "137",
    year = "2012"

}

@article{Sumit:2025ddb,
    author = "Sumit and Parkash, Jai and Das, Santosh K. and Haque, Najmul",
    title = "{Anisotropy effects on heavy quark dynamics in Gribov modified gluon plasma}",
    eprint = "2506.01922",
    archivePrefix = "arXiv",
    primaryClass = "hep-ph",
    month = "6",
    year = "2025"
}

@article{lim2009modeling,
  title={Modeling single-file diffusion with step fractional Brownian motion and a generalized fractional Langevin equation},
  author={Lim, SC and Teo, LP},
  journal={Journal of Statistical Mechanics: Theory and Experiment},
  volume={2009},
  number={08},
  pages={P08015},
  year={2009},
  publisher={IOP Publishing}
}

@article{guo2013numerics,
  title={Numerics for the fractional Langevin equation driven by the fractional Brownian motion},
  author={Guo, Peng and Zeng, Caibin and Li, Changpin and Chen, YangQuan},
  journal={Fractional Calculus and Applied Analysis},
  volume={16},
  number={1},
  pages={123--141},
  year={2013},
  publisher={Versita}
}

@article{10.1111/j.1365-246X.1967.tb02303.x, author = {Caputo, Michele}, title = "{Linear Models of Dissipation whose Q is almost Frequency Independent—II}", journal = {Geophysical Journal International}, volume = {13}, number = {5}, pages = {529-539}, year = {1967}, month = {11}, issn = {0956-540X}, doi = {10.1111/j.1365-246X.1967.tb02303.x},  }

@article{carpinteri2014fractals,
  author = "Carpinteri, Alberto and Mainardi, Francesco",
  title = "{Fractals and Fractional Calculus in Continuum Mechanics}",
  journal = "Fractals and Fractional Calculus in Continuum Mechanics",  
  volume = "378",
  year = "2014",
}

@article{podlubnyacademic,
  title={Academic Press; San Diego: 1999},
  author={Podlubny, I},
  journal={Fractional Differential Equations, Aca-
demic Press, San Diego, 1999}
}

@article{CaputoMainardi1971,
  author  = {M. Caputo and F. Mainardi},
  title   = {Linear Models of Dissipation in Anelastic Solids},
  journal = {Rivista del Nuovo Cimento},
  volume  = {1},
  number  = {2},
  pages   = {161--198},
  year    = {1971},
  doi     = {10.1007/BF02820620}
}

@article{PhysRevC.86.034905,
  title = {MARTINI event generator for heavy quarks: Initialization, parton evolution, and hadronization},
  author = {Young, Clint and Schenke, Bj\"orn and Jeon, Sangyong and Gale, Charles},
  journal = {Phys. Rev. C},
  volume = {86},
  issue = {3},
  pages = {034905},
  numpages = {7},
  year = {2012},
  month = {Sep},
  publisher = {American Physical Society},
  doi = {10.1103/PhysRevC.86.034905},
  url = {https://link.aps.org/doi/10.1103/PhysRevC.86.034905}
}

@article{PhysRevC.93.014901,
  title = {Heavy quark transport in heavy ion collisions at energies available at the BNL Relativistic Heavy Ion Collider and at the CERN Large Hadron Collider within the UrQMD hybrid model},
  author = {Lang, Thomas and van Hees, Hendrik and Inghirami, Gabriele and Steinheimer, Jan and Bleicher, Marcus},
  journal = {Phys. Rev. C},
  volume = {93},
  issue = {1},
  pages = {014901},
  numpages = {16},
  year = {2016},
  month = {Jan},
  publisher = {American Physical Society},
  doi = {10.1103/PhysRevC.93.014901},
  url = {https://link.aps.org/doi/10.1103/PhysRevC.93.014901}
}

@article{PhysRevC.84.064902,
  title = {Thermalization of charm quarks in infinite and finite quark-gluon plasma matter},
  author = {Cao, Shanshan and Bass, Steffen A.},
  journal = {Phys. Rev. C},
  volume = {84},
  issue = {6},
  pages = {064902},
  numpages = {9},
  year = {2011},
  month = {Dec},
  publisher = {American Physical Society},
  doi = {10.1103/PhysRevC.84.064902},
  url = {https://link.aps.org/doi/10.1103/PhysRevC.84.064902}
}

@article{He:2022ywp,
    author = "He, Min and van Hees, Hendrik and Rapp, Ralf",
    title = "{Heavy-quark diffusion in the quark\textendash{}gluon plasma}",
    eprint = "2204.09299",
    archivePrefix = "arXiv",
    primaryClass = "hep-ph",
    doi = "10.1016/j.ppnp.2023.104020",
    journal = "Prog. Part. Nucl. Phys.",
    volume = "130",
    pages = "104020",
    year = "2023"
}

@article{Bhattacharyya:2024hku,
    author = "Bhattacharyya, Trambak and Megias, Eugenio and Deppman, Airton",
    title = "{Jet quenching of the heavy quarks in the quark-gluon plasma and the nonadditive statistics}",
    eprint = "2405.15679",
    archivePrefix = "arXiv",
    primaryClass = "hep-ph",
    doi = "10.1016/j.physletb.2024.138907",
    journal = "Phys. Lett. B",
    volume = "856",
    pages = "138907",
    year = "2024"
}

@article{PhysRevC.84.044901,
  title = {Momentum dependence of drag coefficients and heavy flavor suppression in quark gluon plasma},
  author = {Mazumder, Surasree and Bhattacharyya, Trambak and Alam, Jan-e and Das, Santosh K.},
  journal = {Phys. Rev. C},
  volume = {84},
  issue = {4},
  pages = {044901},
  numpages = {6},
  year = {2011},
  month = {Oct},
  publisher = {American Physical Society},
  doi = {10.1103/PhysRevC.84.044901},
  url = {https://link.aps.org/doi/10.1103/PhysRevC.84.044901}
}

@article{122333
,
    author = "Zhang, Chao and Zheng, Liang and Shi, Shusu and Lin, Zi-Wei",
    title = "{Resolving the RpA and v2 puzzle of D0 mesons in p\ensuremath{-}Pb collisions at the LHC}",
    eprint = "2210.07767",
    archivePrefix = "arXiv",
    primaryClass = "nucl-th",
    doi = "10.1016/j.physletb.2023.138219",
    journal = "Phys. Lett. B",
    volume = "846",
    pages = "138219",
    year = "2023"
}

@article{PhysRevD.103.054030,
  title = {Energy loss versus energy gain of heavy quarks in a hot medium},
  author = {Jamal, Mohammad Yousuf and Das, Santosh K. and Ruggieri, Marco},
  journal = {Phys. Rev. D},
  volume = {103},
  issue = {5},
  pages = {054030},
  numpages = {8},
  year = {2021},
  month = {Mar},
  publisher = {American Physical Society},
  doi = {10.1103/PhysRevD.103.054030},
  url = {https://link.aps.org/doi/10.1103/PhysRevD.103.054030}
}

@article{Jamal:2020emj,
    author = "Jamal, M. Yousuf and Mohanty, Bedangadas",
    title = "{Energy-loss of heavy quarks in the isotropic collisional hot QCD medium at a finite chemical potential}",
    eprint = "2002.09230",
    archivePrefix = "arXiv",
    primaryClass = "nucl-th",
    doi = "10.1140/epjp/s13360-021-01098-4",
    journal = "Eur. Phys. J. Plus",
    volume = "136",
    number = "1",
    pages = "130",
    year = "2021"
}

@article{rogers1997arbitrage,
  title={Arbitrage with fractional Brownian motion},
  author={Rogers, L Chris G},
  journal={Mathematical finance},
  volume={7},
  number={1},
  pages={95--105},
  year={1997},
  publisher={Wiley Online Library}
}

@article{Zhang:2022fum,
    author = "Zhang, Chao and Zheng, Liang and Shi, Shusu and Lin, Zi-Wei",
    title = "{Resolving the RpA and v2 puzzle of D0 mesons in p\ensuremath{-}Pb collisions at the LHC}",
    eprint = "2210.07767",
    archivePrefix = "arXiv",
    primaryClass = "nucl-th",
    doi = "10.1016/j.physletb.2023.138219",
    journal = "Phys. Lett. B",
    volume = "846",
    pages = "138219",
    year = "2023"
}

@article{Sun:2023adv,
    author = "Sun, Yifeng and Plumari, Salvatore and Das, Santosh K.",
    title = "{Exploring the effects of electromagnetic fields and tilted bulk distribution on directed flow of D mesons in small systems}",
    eprint = "2304.12792",
    archivePrefix = "arXiv",
    primaryClass = "nucl-th",
    doi = "10.1016/j.physletb.2023.138043",
    journal = "Phys. Lett. B",
    volume = "843",
    pages = "138043",
    year = "2023"
}

@article{Plumari:2019hzp,
    author = "Plumari, Salvatore and Coci, Gabriele and Minissale, Vincenzo and Das, Santosh K. and Sun, Yifeng and Greco, Vincenzo",
    title = "{Heavy - light flavor correlations of anisotropic flows at LHC energies within event-by-event transport approach}",
    eprint = "1912.09350",
    archivePrefix = "arXiv",
    primaryClass = "hep-ph",
    doi = "10.1016/j.physletb.2020.135460",
    journal = "Phys. Lett. B",
    volume = "805",
    pages = "135460",
    year = "2020"
}

@article{Prakash_2024,
doi = {10.1088/1361-6471/ad10c9},
url = {https://dx.doi.org/10.1088/1361-6471/ad10c9},
year = {2023},
month = {dec},
publisher = {IOP Publishing},
volume = {51},
number = {2},
pages = {025101},
author = {Jai Prakash and Mohammad Yousuf Jamal},
title = {Heavy quark energy loss through polarization and radiation in the hot QCD medium},
journal = {Journal of Physics G: Nuclear and Particle Physics},
}

@article{du2023accelerated,
  title={Accelerated quantum circuit Monte-Carlo simulation for heavy quark thermalization},
  author={Du, Xiaojian and Qian, Wenyang},
  journal={arXiv preprint arXiv:2312.16294},
  year={2023}
}

@article{Prakash:2023hfj,
    author = "Prakash, Jai and Jamal, Mohammad Yousuf",
    title = "{Study of the heavy quarks energy loss through medium polarization, elastic collision and radiative processes}",
    eprint = "2304.04003",
    archivePrefix = "arXiv",
    primaryClass = "nucl-th",
    month = "4",
    year = "2023"
}

@article{Shaikh:2021lka,
    author = "Shaikh, Adiba and Kurian, Manu and Das, Santosh K. and Chandra, Vinod and Dash, Sadhana and Nandi, Basanta K.",
    title = "{Heavy quark transport coefficients in a viscous QCD medium with collisional and radiative processes}",
    eprint = "2105.14296",
    archivePrefix = "arXiv",
    primaryClass = "hep-ph",
    doi = "10.1103/PhysRevD.104.034017",
    journal = "Phys. Rev. D",
    volume = "104",
    number = "3",
    pages = "034017",
    year = "2021"
}

@article{He:2013zua,
    author = "He, Min and van Hees, Hendrik and Gossiaux, Pol B. and Fries, Rainer J. and Rapp, Ralf",
    title = "{Relativistic Langevin Dynamics in Expanding Media}",
    eprint = "1305.1425",
    archivePrefix = "arXiv",
    primaryClass = "nucl-th",
    doi = "10.1103/PhysRevE.88.032138",
    journal = "Phys. Rev. E",
    volume = "88",
    pages = "032138",
    year = "2013"
}

@article{He:2012df,
    author = "He, Min and Fries, Rainer J. and Rapp, Ralf",
    title = "{$\mathbf{D_s}$-Meson as Quantitative Probe of Diffusion and Hadronization in Nuclear Collisions}",
    eprint = "1204.4442",
    archivePrefix = "arXiv",
    primaryClass = "nucl-th",
    doi = "10.1103/PhysRevLett.110.112301",
    journal = "Phys. Rev. Lett.",
    volume = "110",
    number = "11",
    pages = "112301",
    year = "2013"
}

@ARTICLE{2014JMP....55b3301S,
       author = {{Sandev}, Trifce and {Metzler}, Ralf and {Tomovski}, {\v{Z}}ivorad},
        title = "{Correlation functions for the fractional generalized Langevin equation in the presence of internal and external noise}",
      journal = {Journal of Mathematical Physics},
         year = 2014,
        month = feb,
       volume = {55},
       number = {2},
          eid = {023301},
        pages = {023301},
          doi = {10.1063/1.4863478},
       adsurl = {https://ui.adsabs.harvard.edu/abs/2014JMP....55b3301S}
}

@article{Kapusta:2014dja,
    author = "Kapusta, J. I. and Young, C.",
    title = "{Causal Baryon Diffusion and Colored Noise}",
    eprint = "1404.4894",
    archivePrefix = "arXiv",
    primaryClass = "nucl-th",
    doi = "10.1103/PhysRevC.90.044902",
    journal = "Phys. Rev. C",
    volume = "90",
    number = "4",
    pages = "044902",
    year = "2014"
}

@article{ WOS:000332486500022,
Author = {Sandev, Trifce and Metzler, Ralf and Tomovski, Zivorad},
Title = {Correlation functions for the fractional generalized Langevin equation
   in the presence of internal and external noise},
Journal = {JOURNAL OF MATHEMATICAL PHYSICS},
Year = {2014},
Volume = {55},
Number = {2},
Month = {FEB},
DOI = {10.1063/1.4863478},
Article-Number = {023301},
ISSN = {0022-2488},
EISSN = {1089-7658},
ResearcherID-Numbers = {Metzler, Ralf/J-9088-2013
   Tomovski, Zivorad/AAJ-3329-2020
   Sandev, Trifce/A-1793-2011},
ORCID-Numbers = {Metzler, Ralf/0000-0002-6013-7020
   Sandev, Trifce/0000-0001-9120-3847},
Unique-ID = {WOS:000332486500022},
}

@article{ WOS:000291660200006,
Author = {Kneller, Gerald R.},
Title = {Generalized Kubo relations and conditions for anomalous diffusion:
   Physical insights from a mathematical theorem},
Journal = {JOURNAL OF CHEMICAL PHYSICS},
Year = {2011},
Volume = {134},
Number = {22},
Month = {JUN 14},
DOI = {10.1063/1.3598483},
Article-Number = {224106},
ISSN = {0021-9606},
ORCID-Numbers = {Kneller, Gerald/0000-0002-3374-3797},
Unique-ID = {WOS:000291660200006},
}

@article{Ciesielski2003,
  author    = {M. Ciesielski and J. Leszczyński},
  title     = {Anomalous diffusion in the Caputo fractional derivative framework},
  journal   = {The European Physical Journal B},
  volume    = {36},
  year      = {2003},
  pages     = {33--35},
  doi       = {10.1140/epjb/e2003-00308-4}
}

@article{Sandev2015,
  author    = {T. Sandev and A. Iomin and H. Kantz and R. Metzler},
  title     = {Langevin equations for a class of Lévy walk processes},
  journal   = {Journal of Physics A: Mathematical and Theoretical},
  volume    = {48},
  number    = {39},
  year      = {2015},
  pages     = {395002},
  doi       = {10.1088/1751-8113/48/39/395002}
}

@article{Fa2007,
  author    = {K. Fa and E. Lenzi},
  title     = {Generalized Langevin equation with memory kernel based on fractional derivatives},
  journal   = {Physical Review E},
  volume    = {75},
  year      = {2007},
  pages     = {061118},
  doi       = {10.1103/PhysRevE.75.061118}
}
\end{document}